\pdfoutput=1
\documentclass[
10pt,
amsmath,amssymb,aps,prx,superscriptaddress,
noeprint,
floatfix,
twocolumn
]{revtex4-1}
\usepackage[utf8]{inputenc}
\usepackage[T1]{fontenc}
\usepackage{microtype}
\usepackage{amsthm}
\usepackage{graphicx}
\usepackage{dcolumn} 
\usepackage{bm} 
\usepackage{bbm} 
\usepackage{mathtools}
\usepackage{algorithm}
\usepackage{algpseudocode}
\usepackage{xparse} 
\usepackage[dvipsnames]{xcolor}

\usepackage{tikz}
\usepackage[braket]{qcircuit}
\usetikzlibrary{cd}
\usetikzlibrary{positioning}
\usetikzlibrary{shapes,arrows}

\usepackage{braket}

\usepackage[colorlinks=true,bookmarks=false,linkcolor=NavyBlue,urlcolor=NavyBlue,citecolor=NavyBlue,breaklinks]{hyperref}

\graphicspath{{figures/}}

\newcommand{\be}{\begin{equation}}
\newcommand{\ee}{\end{equation}}
\newcommand{\ben}{\begin{eqnarray}}
\newcommand{\een}{\end{eqnarray}}
\newcommand{\bes}{\begin{subequations}}
\newcommand{\ees}{\end{subequations}}

\newcommand{\id}{\mathbbm{1}}

\newcommand{\jsin}{(J^S)^{-1}}

\newcommand{\C}{\mathbb{C}}
\newcommand{\I}{\mathrm{i}}

\newcommand{\rofp}{r_{1,4}^{+}}
\newcommand{\rofm}{r_{1,4}^{-}}
\newcommand{\rttp}{r_{2,3}^{+}}

\newcommand{\proj}[1]{\left |{#1} \left \rangle \! \right \langle {#1}\right |}
\let\Re\relax
\let\Im\relax
\DeclareMathOperator{\Re}{Re}
\DeclareMathOperator{\Im}{Im}

\DeclareMathOperator{\Tr}{Tr}
\DeclareMathOperator{\tr}{Tr}

\DeclareMathOperator{\spn}{span}

\DeclareMathOperator*{\argmin}{arg\,min}
\renewcommand{\vec}{\boldsymbol}

\newcommand{\CH}{C^{\textrm{H}}}
\newcommand{\CC}{C^{\textrm{C}}}

\DeclarePairedDelimiter{\parens}{(}{)}
\NewDocumentCommand{\of}{m}{\parens*{#1}}
\DeclarePairedDelimiter{\brackets}{[}{]}
\NewDocumentCommand{\off}{m}{\brackets*{#1}}

\tikzstyle{decision} = [diamond, draw, fill=blue!20, 
    text width=4.5em, text badly centered, node distance=3cm, inner sep=0pt]
\tikzstyle{block} = [rectangle, draw, fill=blue!20, 
    text width=5em, text centered, rounded corners, minimum height=4em]
\tikzstyle{line} = [draw, -latex']
\tikzstyle{cloud} = [draw, ellipse,fill=red!20, node distance=3cm,
    minimum height=2em]

\begin{document}
	
\title{Attainability of the Holevo-Cramér-Rao bound for two-qubit 3D magnetometry}

\author{Jamie Friel}
\email{j.friel@warwick.ac.uk}
\affiliation{Department of Physics, University of Warwick, Coventry CV4 7AL, United Kingdom}
\affiliation{EPSRC Centre for Doctoral Training in Diamond Science and Technology, UK}

\author{Pantita Palittapongarnpim}
\email{panpalitta@gmail.com}
\affiliation{Department of Physics, University of Warwick, Coventry CV4 7AL, United Kingdom}

\author{Francesco Albarelli}
\email{francesco.albarelli@gmail.com}
\affiliation{Faculty of Physics, University of Warsaw, 02-093 Warszawa, Poland}

\author{Animesh Datta}
\email{Animesh.Datta@warwick.ac.uk}
\affiliation{Department of Physics, University of Warwick, Coventry CV4 7AL, United Kingdom}
\date{\today}

\begin{abstract}

	We study quantum-limited 3D magnetometry using two qubits.
	Two qubits form the smallest multi-qubit system for 3D magnetometry, the simultaneous estimation of three phases, as it is impossible with a single qubit.
	We provide an analytical expression for the Holevo-Cramér-Rao bound (HCRB), the fundamental attainable quantum bound of multiparameter estimation, for 3D magnetometry using two-qubit pure states and show its attainability by rank-1 projective measurements.
	We also examine the attainability of the HCRB in the presence of dephasing noise using numerical methods.
	While attaining the HCRB may require collective measurements over infinitely many copies, we find that for high noise the HCRB is practically saturated by two copies only.
	In the low noise regime, up to three copies are unable to attain the HCRB.
	More generally, we introduce new multiparameter \emph{channel bounds} to compare quantum-classical and classical-quantum strategies where multiple independent copies of the state are entangled before or after recording the parameters respectively.
	We find that their relative performance depends on the noise strength, with the classical-quantum strategy performing better for high noise.
	We end with shallow quantum circuits that approach the fundamental quantum limit set by the HCRB for two-qubit 3D magnetometry using up to three copies.
\end{abstract}

\maketitle
	
\section{\label{sec:itnro} Introduction}

The study of how to perform very accurate and precise measurements of physical quantities using quantum probes has been an extremely lively and fruitful line of research~\cite{Giovannetti2011,Demkowicz-Dobrzanski2015a,Degen2016,Pezze2018,Polino2020}.
A prominent application is magnetometry where measurement precision has dramatically improved in recent years via the use of quantum probes~\cite{Sewell2012,Dale2017,Casola2018,Barry2019}.
In particular, the problem of how to optimally estimate several parameters of interest \emph{simultaneously} has become topical in recent years~\cite{Szczykulska2016,Liu2019d,Albarelli2019c,Demkowicz-Dobrzanski2020}.
Estimating multiple parameters simultaneously is important in several high-level applications such as imaging which involves many pixels, spectroscopy which involves many frequencies, accelerometry and magnetometry which involves three-dimensional (3D) fields. In addition, signals in these applications often vary across space and time. While multiple parameters may be sensed sequentially, it may be too slow and certain parameters may vary while others are being estimated. Independent sensors may be used to circumvent this issue, but add to the footprint and could lead to conflicts in reconciling their estimates~\cite{Datta2020}.

More fundamentally, simultaneous quantum-limited estimation of multiple parameters provides the rightful landscape for studying incompatibility in quantum mechanics in a quantitative manner, most rigorously and aesthetically in the language of non-commutative information geometry~\cite{Hayashi2005}.
Known colloquially as multiparameter quantum estimation theory, its objective is to identify the fundamental bounds that quantum mechanics imposes on the precisions with which parameters encoded in quantum states can be estimated and optimal strategies for attaining them.
A lower bound on the variance of an unbiased estimator~\cite{helstrom1976quantum} is imposed by the quantum Cramér-Rao bound (CRB). It is given by the inverse of the quantum Fisher information (QFI) and is always saturable for single parameters~\cite{Braunstein1994,Barndorff-Nielsen2000,Hayashi2005}.
Its naive multiparameter extension that relies on the QFI matrix, however, is not guaranteed to be attainable.
The fundamental \emph{attainable} bound for multiparameter quantum estimation instead is the Holevo-Cramér-Rao bound (HCRB)~\cite{Holevo1976,Holevo2011b,Demkowicz-Dobrzanski2020}.
Although the HCRB and the scalar CRB obtained from the QFI matrix differ by no more than a factor of two~\cite{Carollo2019,Tsang2019}, the former provides insights into the optimal measurements for attainability, as we will show.

In this paper we focus on quantum-limited 3D magnetometry, the estimation of the three components of a magnetic field acting on a quantum system with precision solely limited by quantum mechanics.
Mathematically, this amounts to estimating three phase parameters, encoded by noncommuting generators, simultaneously.
As this is impossible with a single qubit,  we concentrate on the smallest multi-qubit system, i.e., two qubits, that allows sensing all three components simultaneously.
Similar problems have been studied by considering the magnetic field acting on a single system with an additional ancillary system unaffected by the dynamics~\cite{DAriano2001c,Fujiwara2001a,Acin2001a,Ballester2004,Chiribella2005,Imai2007,Yuan2016b,Kura2017}; in such a scenario it is sufficient to consider the QFI matrix.
On the contrary, we do not consider ancillary systems in this work.
This scenario has been studied in the QFI matrix formalism~\cite{Kolenderski2008,Baumgratz2015,Ho2020} and applied to photonic systems~\cite{Goldberg2018,Liu2017h}.
However, the fundamental attainable bound for 3D magnetometry remains unknown.

In this paper, we first present a closed-form analytical expression for the fundamental attainable bound---the HCRB---for 3D magnetometry for two-qubit pure states and show that it can be attained by a rank-1 projective measurement.
We then study the impact of noise, focusing on independent and identical dephasing  in the $z$ direction on the two qubits.
Unlike pure states, the HCRB for mixed states is, in principle, attained asymptotically in the number of identical copies of the state, which must be measured collectively~\cite{Guta2007,Demkowicz-Dobrzanski2020}. 
As this is prohibitive in practice we study the contribution of a few (two and three) copies of the two-qubit state towards reaching the asymptotic limit, by numerical optimisation of the classical CRB over collective measurements.
Our second result shows that this contribution is related to the strength of the noise.

When using a noisy quantum channel repeatedly for transmitting classical or quantum information~\cite{Wilde2011}, it is possible to employ quantum entanglement only at the preparation stage in a quantum-classical (QC) strategy, only at the measurement stage in a classical-quantum (CQ) strategy, or at both ends in a quantum-quantum (QQ) strategy.
A similar approach to quantum estimation strategies shows that for single parameter estimation CQ strategies are never useful~\cite{Giovannetti2006,Hayashi2005}. 
The same is not true for multiple parameters, as exemplified by the fact that collective measurements are needed to attain the HCRB.
Thus, we investigate the performance of a QC strategy by numerically optimising the classical CRB over initial $2 k$-qubit states and over $k$ independent measurements over 2 qubits systems.
Our third result is that for $k=2$ in the high noise regime, the CQ strategy (which approaches the HCRB for high noise) outperforms the QC strategy.

Moving further towards practical implementation, we present numerically-optimised, shallow quantum circuits executing CQ strategies for quantum-limited
two-qubit 3D magnetometry with up to three copies using up to six qubits.
The circuits were optimised independently from the previous numerical optimisation of the classical CRB over collective measurements. That they both provide commensurate results provides additional reassurance on the validity of our non-analytical results.

To help the reader navigate this paper, we summarise our main results ranging from the fundamental to practical quantum-limited two-qubit 3D magnetometry.
\begin{enumerate}
	\item 
\begin{enumerate}
	\item A closed-form expression for the noiseless HCRB~(Eq.~\eqref{equ:ch}) (for a vanishing magnetic field)
	\item A proof of attainability of the noiseless HCRB with standard (rank-1) projective measurements in Sec.~\ref{subsec:projattain}
\end{enumerate}	
	\item In the presence of dephasing, a numerical comparison between strategies with $k=2,3$ copies and the optimal HCRB in Fig.~\ref{FIG:opt_state_and_measure}, showing that the attainability of the HCRB depends on the noise strength.
	\item Also in Fig.~\ref{FIG:opt_state_and_measure}, a numerical comparison between CQ and QC strategies for $k=2$ copies, showing that for high noise the CQ strategy gives a smaller estimation error.
\end{enumerate}

The manuscript is structured as follows.
In Sec.~\ref{sec:metrol} we briefly introduce multiparameter quantum estimation theory and define a new class of multiparameter \emph{channel bounds} for a finite number of copies.
In Sec.~\ref{sec:analytics} we analytically calculate the HCRB explicitly for 3D magnetometry and show that this bound is attainable with a projective measurement.
In Sec.~\ref{sec:noisy} we study the contribution of additional identical copies of the state on the attainability of the HCRB when dephasing noise is present.
In Sec.~\ref{sec:circ} we discuss practical implementations in form of shallow quantum circuits.
In Sec.~\ref{sec:concl} we close by discussing some open problems.

\section{\label{sec:metrol} Multiparameter estimation}

The goal of multiparameter quantum metrology is to estimate multiple parameters encoded into some initial state by an external process optimally within the laws of quantum mechanics.
There is, in general, no unique notion of optimality and we choose to consider the sum of variances of all parameters as the scalar figure of merit.

More formally, given a finite-dimensional Hilbert space $\mathcal{H}$, we consider a family of quantum states $\rho_{\vec{\varphi}}$ that depends on a vector $\vec{\varphi}$ of $p$ real parameters. The set $\{\vec{\varphi}, \rho_{\vec{\varphi}}, \mathcal{H} \}$ is also known as a quantum statistical model.
Estimation is then performed from the measurement outcomes $x$, with probability given by the Born rule $p(x|\vec{\varphi}) = \Tr[\rho_{\vec{\varphi}} \Pi_x]$, where $\Pi_x$ is an element of a positive, operator-valued measure (POVM) $\Pi = \left\lbrace\Pi_x \ge 0 | \sum_x \Pi_x = \id \right\rbrace$ that describes the statistics of the measurement apparatus.
In particular, we often focus on rank-1 projective measurements $\Pi_x = | x \rangle \langle x |$, where $\ket{x}$ is an orthonormal basis; for brevity we will refer to them as projective measurements.
The function of the outcomes $x$ that gives an estimate $\tilde{\vec{\varphi}}(x)$ is the estimator and its precision can be quantified by the mean square error matrix (MSEM) 
\begin{equation}
\label{eq:MSEM}
V_{\vec{\varphi}} (\Pi, \Tilde{\vec{\varphi}})_{i,j} = \sum_x p(x|\vec{\varphi})[\Tilde{\varphi}(x)_i-\varphi_i][\Tilde{\varphi}(x)_j-\varphi_j].
\end{equation}
To compare estimation errors in a strictly ordered way we define the scalar figure of merit~\footnote{More generally it is common to consider the figure of merit $\Tr W V_{\vec{\varphi}}$ depending on a positive weight matrix $W$; however, in this work we choose $W = \id$, except in Appendix \ref{app:weight_maxt} where weight matrices are considered. },
\begin{equation}
\Delta ^2\Tilde{\vec{\varphi}} = \Tr  \left[V_{\vec{\varphi}} (\Pi, \Tilde{\vec{\varphi}}) \right].
\end{equation}
Furthermore, we restrict ourselves to locally unbiased estimators that satisfy for all $i,j$~\cite{Holevo2011b}
\begin{align}
\sum_x p(x|\vec{\varphi})(\Tilde{\varphi}(x)_i-\varphi_i) &= 0,\\
\sum_x \Tilde{\varphi}(x)_i \frac{\partial p(x|\vec{\varphi})}{\partial \varphi_j} &= \delta_{ij}.
\end{align}
Here and henceforth, all quantities are evaluated at the true value of the parameter.
The MSEM in Eq.~\eqref{eq:MSEM} for such estimators is just the covariance matrix, which satisfies the lower bound $V_{\vec{\varphi}} (\Pi, \Tilde{\vec{\varphi}}) \geq F(\rho_{\vec{\varphi}},\Pi)^{-1}$, in terms of the classical Fisher information (CFI) matrix~\cite{Liu2019d}
\begin{equation}
\label{eq:CFIM}
F(\rho_{\vec{\varphi}},\Pi)_{ij} = \sum_{x} p( x | \vec{\varphi} ) \left( \frac{ \partial \log p( x | \vec{\varphi} ) }{\partial \varphi_i} \right) \left( \frac{ \partial \log p( x | \vec{\varphi} ) }{\partial \varphi_j} \right).
\end{equation}
We will mainly focus on the corresponding \emph{scalar} bound $\CC(\rho_{\vec{\varphi}},\Pi)$ such that
\begin{equation}
\label{eq:classCRB}
\Delta ^2\Tilde{\vec{\varphi}} \geq \Tr [ F(\rho_{\vec{\varphi}},\Pi)^{-1} ] = \CC(\rho_{\vec{\varphi}},\Pi),
\end{equation}
where the superscript stands for classical, in contrast to quantum bounds which will depend only on the family $\rho_{\vec{\varphi}}$ and not on the measurement.

The CFI matrix is upper bounded as $F(\rho_{\vec{\varphi}},\Pi) \leq J(\rho_{\vec{\varphi}}) $ by the QFI matrix
\begin{equation}
\label{eq:QFImat}
J(\rho_{\vec{\varphi}})_{ij} =  \Tr \left[ \frac{ L_i L_j + L_j L_i }{2}  \rho_{\vec{\varphi}} \right],
\end{equation}
defined in terms of the symmetric logarithmic derivatives (SLDs) $\{ L_i \}$ satisfying $\partial \rho_{\vec{\varphi}}/\partial \varphi_i = (L_i \rho_{\vec{\varphi}}  + \rho_{\vec{\varphi}}  L_i )/2 $ and $L_i^\dag = L_i$.
The corresponding scalar bound $C^S (\rho_{\vec{\varphi}})$ is~\footnote{The superscript $S$ denotes that the bound is obtained from the SLDs rather than other  logarithmic derivatives.  We will not use them in this work as they are less informative than the HCRB.}
\begin{equation}
\Delta ^2\Tilde{\vec{\varphi}} \geq \CC(\rho_{\vec{\varphi}},\Pi) \geq C^S (\rho_{\vec{\varphi}}) = \Tr \left[J(\rho_{\vec{\varphi}})^{-1}\right].
\end{equation}

The Holevo-Cramér-Rao bound (HCRB) introduced by Holevo~\cite{Holevo2011b}, in the equivalent formulation given by Nagaoka~\cite{Nagaoka1989}, is defined as
\begin{subequations}
\label{eq:HCRB}
\begin{equation}
\CH \of{ \rho_{\vec{\varphi}} } =  \min_{X } \tr \Re Z[X] + \left\lVert \Im Z[X] \right\rVert_1 ,
\end{equation}    
\begin{equation}
\text{s.t.} \quad \Tr X_i \frac{\partial \rho_{\vec{\varphi}}}{\partial \varphi_j} = \delta_{ij},  \label{eq:HCRBconst} 
\end{equation}    
\begin{equation} Z[X]_{ij} = \Tr[\rho_{\vec{\varphi}} X_i X_j] \label{eq:HCRBZmat}
\end{equation}
\end{subequations}
where $X = \{ X_1 , \dots  X_p \}$ is a collection of $p$ Hermitian matrices, $\Re$ and $\Im$ denote the elementwise real and imaginary part of a matrix and $\left\lVert \cdot \right\rVert_1$ denotes the trace norm (sum of the singular values).
Crucially, this bound is always tighter than the SLD one. Mathematically,
\begin{equation} 
\Delta ^2\Tilde{{\vec{\varphi}}} \geq \CC(\rho_{\vec{\varphi}},\Pi) \geq \CH(\rho_{\vec{\varphi}})  \geq  C^S (\rho_{\vec{\varphi}}),
\end{equation}
meaning that it takes better account of the possible incompatibility of the optimal observables while estimating multiple parameters simultaneously.
It is important to note that the HCRB often collapses to other scalar quantum CRBs~\cite{Albarelli2019c,Suzuki2018}.
The constrained optimisation~\eqref{eq:HCRB} has recently been shown to be a convex problem~\cite{Albarelli2019} and can be rewritten as a semidefinite program, which drastically reduces the cost of numerical evaluation.

The minimisation~\eqref{eq:HCRB} does not guarantee the existence of a POVM acting on $\mathcal{H}$ such that $\CC(\rho_{\vec{\varphi}},\Pi) = \CH(\rho_{\vec{\varphi}})$, apart from particular cases such as pure states~\cite{Matsumoto2002}.
For mixed states, the HCRB is also attainable~\cite{Hayashi2008a,Kahn2009,Yamagata2013,Yang2018a,Demkowicz-Dobrzanski2020}, although  requiring, in general, a collective measurement on an asymptotically large number of identical copies of $\rho_{\vec{\varphi}}$.

This gives another definition of the HCRB as
\begin{equation}
\label{eq:CHlim}
\CH \of{\rho_{\vec{\varphi}}} = \lim_{k \rightarrow \infty}  \min_{\Pi^{(k)}} k \, \CC(\rho_{\vec{\varphi}}^{\otimes k},\Pi^{(k)}),
\end{equation}
where $\Pi^{(n)}$ is a POVM acting on the Hilbert space $\mathcal{H}^{\otimes n}$.
We further introduce a class of $k$-copies attainable bounds~\footnote{This bound for $k=1$ is also known as the  \emph{most informative} bound~\cite{Nagaoka1989}, not to be confused with the maximum between SLD and RLD bounds~\cite{Genoni2013b}.}
\begin{equation} 
\label{eq:kcopyCRB}
C^{(k)}\of{\rho_{\vec{\varphi}}}= \min_{\Pi^{(k)}} k \, \CC \of{\rho_{\vec{\varphi}}^{\otimes k},\Pi^{(k)} },
\end{equation}
so that formally we have $C^{(\infty)}\of{\rho_{\vec{\varphi}}} = \CH\of{\rho_{\vec{\varphi}}}$.

We stress that increasing the number of identical copies of the quantum state does not correspond to identical and independently distributed classical random variables describing the classical outcomes, unless the measurements are performed on each copy independently.
It follows that there are two different kind of asymptotics that enter into the saturation of multiparameter quantum CRBs---one in the number of identical copies and one in the number of repetitions of the experiment, necessary in general to saturate the classical CRB~\eqref{eq:classCRB} with a classical estimator~\footnote{In the asymptotic limit this is issue is not relevant, because of the Gaussian nature of the asymptotic quantum statistical model~\cite{Guta2007,Hayashi2008a,Kahn2009,Yamagata2013,Yang2018a,Demkowicz-Dobrzanski2020}.}.
In this paper we focus only on the former as it is a uniquely quantum aspect of multiparameter estimation.
Previous works have shown the usefulness of entangled measurements over multiple copies for multiparameter estimation, both theoretically~\cite{Gill2000,Bagan2006a,Vidrighin2014,Zhu2018} and experimentally~\cite{Roccia2017,Parniak2018,Hou2018}.
Entangled measurements are also useful for noisy single-parameter estimation, but in that case the entanglement is between the partitions of a single copy of the system~\cite{Micadei2015,Piera2020} or with ancillas~\cite{Demkowicz-Dobrzanski2014,Huang2016a,Wang2018b,Sbroscia2018}, not between identical copies.
\subsection{Pure states}
\label{sec:ps}

For pure states $\rho_{\vec{\varphi}} = | \psi_{\vec{\varphi}} \rangle \langle \psi_{\vec{\varphi}} | $ it is possible to recast the optimisation~\eqref{eq:HCRB} in terms of complex vectors~\cite{Matsumoto2002,Hayashi2017c}
\begin{equation}
\ket{x_i}=  X_i \ket{\psi_{\vec{\varphi}}},
\end{equation}
for which the matrix~\eqref{eq:HCRBZmat} becomes 
$Z[X]_{ij} = \braket{x_i|x_j}$
and the constraints~\eqref{eq:HCRBconst} become
\begin{equation}
\label{eq:HCRBpureconst}
\Re \braket{x_i|l_j} = \delta_{ij},
\end{equation}
where we have introduced the vectors
\begin{equation}
\label{eq:horlift}
\ket{l_i} = L_i \ket{\psi} = 2 \left( \ket{\partial_i \psi_{\vec{\varphi}} } - \braket{\psi_{\vec{\varphi}} | \partial_i \psi_{\vec{\varphi}} }\ket{\psi_{\vec{\varphi}}} \right).
\end{equation}
A crucial simplification is that the vectors $\ket{x_i}$ attaining the minimum can be always found in $\spn_{\C} \left\{ \ket{l_i} \right\}_{i=1}^p$~\cite{Matsumoto2002}.
Furthermore, as we have already mentioned, for pure states the limit in Eq.~\eqref{eq:CHlim} is not needed and measurements on single copies of the systems attain the HCRB, i.e., $C^{(1)}\of{| \psi_{\vec{\varphi}} \rangle \langle \psi_{\vec{\varphi}} | } = \CH \of{ | \psi_{\vec{\varphi}} \rangle \langle \psi_{\vec{\varphi}} | }$.
Since the introduction of the HCRB in 1976~\cite{Holevo1976}, very few analytic solutions to minimisation~\eqref{eq:HCRB} have been found for nontrivial cases~\footnote{In the context of evaluating the HCRB, with the phrase \emph{trivial cases} we refer to those quantum statistical models for which the HCRB collapses to other bounds (SLD or RLD) that are easier to evaluate~\cite{Albarelli2019c,Suzuki2018}.}, mostly for two-parameter problems~\cite{Matsumoto2002,Bradshaw2017,Suzuki2016a,Sidhu2019a}.
In Sec.~\ref{sec:analytics} we present the first analytic expression for the HCRB for a nontrivial three-parameter pure-state model motivated by 3D magentometry.

\subsection{Channel bounds}\label{subsec:chbound}
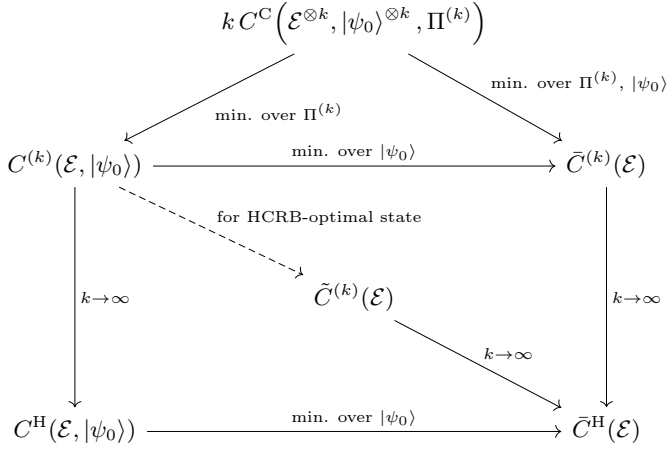
\begin{figure}[t]
\begin{equation*}
\begin{tikzcd}[row sep = huge]
& k \, \CC \of{ \mathcal{E}^{\otimes k},\ket{\psi_0}^{\otimes k},\Pi^{(k)} } \arrow[ld, "\text{min. over } \Pi^{(k)}"]\arrow[rd, "\text{min. over } \Pi^{(k)} \text{, } \ket{\psi_0}"] &  \\
C^{(k)}\of{\mathcal{E},\ket{\psi_0}} \arrow[dd, "k \to \infty"]\arrow[rd, dashed, "\text{for HCRB-optimal state}"] \arrow[rr, "\text{min. over } \ket{\psi_0}"]& & \bar{C}^{(k)}\of{\mathcal{E}}\arrow[dd, "k \to \infty"]  \\
&  \tilde{C}^{(k)}\of{\mathcal{E}} \arrow[rd, "k \to \infty"]  &  \\
\CH\of{\mathcal{E},\ket{\psi_0}} \arrow[rr, "\text{min. over } \ket{\psi_0}"] & &  \bar{C}^{\text{H}}\of{\mathcal{E}}
\end{tikzcd}
\end{equation*}
\caption{
Relations between attainable scalar bounds for multiparameter estimation with $k$ identical copies of the system.
Solid arrows point towards the smaller or equal object in the pair; the only inequality not shown in the diagram is $\tilde{C}^{(k)}(\mathcal{E}) \geq \bar{C}^{(k)}(\mathcal{E})$. All quantities defined via minimisations over POVMs also admit a projective equivalent denoted by a $*$ in the subscript such as $C^{(k)}_*\of{\rho_{\vec{\varphi}}}$ defined in Eq.~\eqref{eq:projkcopyCRB}.
}
\label{fig:crbs-cd}
\end{figure}

It is natural in estimation schemes to separate the preparation of an initial state $\ket{\psi_0}$ from the parameter encoding performed by a quantum channel $\mathcal{E}_{\vec{\varphi}}$, such that $\rho_{\vec{\varphi}} = \mathcal{E}_{\vec{\varphi}}\off{| \psi_0 \rangle \langle \psi_0 | } $.
We will therefore separate the dependence on the encoding and the initial state in the various CRBs, e.g., $\CC\of{\mathcal{E},\ket{\psi_0},\Pi}$ and $C^{(k)}\of{\mathcal{E},\ket{\psi_0}}$.

Having a scalar figure of merit that quantifies the estimation error, it is now natural to define \emph{channel bounds} by minimizing over the initial state.
While general results on channel bounds for single parameter estimation are known~\cite{Fujiwara2001,Sarovar2004,Fujiwara2008,Hayashi2011,Demkowicz-Dobrzanski2012,Katariya2020a}, the multiparameter case is still largely unexplored.
To that end, we define the optimised HCRB
\begin{equation}
\bar{C}^{\textrm{H}}\of{\mathcal{E}} = \min_{\ket{\psi_0}} \CH\of{\mathcal{E},\ket{\psi_0}},
\end{equation}
for which the optimal state is denoted as $ \ket{\Tilde{\psi}_0} = \argmin_{\ket{\psi_0}} \CH\of{\mathcal{E},\ket{\psi_0}}$.
To study how this asymptotic quantity is approached, we introduce both the fully optimised $k$-copy channel bound
\begin{equation}
\bar{C}^{(k)}\of{\mathcal{E}} =  \min_{\ket{\psi_0},\Pi^{(k)}} k \, \CC \of{ \mathcal{E}^{\otimes k},\ket{\psi_0}^{\otimes k},\Pi^{(k)} },
\end{equation}
and the $k$-copy bound~\eqref{eq:kcopyCRB} of the HCRB-optimal initial state
\begin{equation}
\tilde{C}^{(k)}\of{\mathcal{E}} = C^{(k)}\of{\mathcal{E},\ket{\tilde{\psi}_0}},
\end{equation}
for which in general $\tilde{C}^{(k)}\of{\mathcal{E}}  \geq \bar{C}^{(k)}\of{\mathcal{E}}$, even though they both tend to $\bar{C}^{\textrm{H}}\of{\mathcal{E}}$ as $k \to \infty$.
The relationships between the various bound considered in this work is schematically summarised in Fig.~\ref{fig:crbs-cd}.

\section{\label{sec:analytics} HCRB for pure two-qubit states}

Mathematically, 3D magnetometry can be cast as the estimation of 3 dimensionless parameters $\vec{\varphi}=[\varphi_1,\varphi_2,\varphi_3]$ in the single-qubit Hamiltonian $h(\vec{\varphi})=\sum_{i=1}^{3} \varphi_i \sigma_i$, where $\{\sigma_i\}$ are the Pauli matrices and $\vec{\varphi}$ is proportional to the three spatial components of a magnetic field.
The same Hamiltonian acts independently on each qubit, so the unitary acting on each two-qubit state is $U(\vec{\varphi})^{\otimes 2} = e^{-\I h(\vec{\varphi})} \otimes e^{-\I h(\vec{\varphi})}$ (See Appendix~\ref{app:mag_details}).
As mentioned previously, the HCRB often collapses to other statistical bounds, although not for our problem (See Appendix~\ref{app_asym_clas}).

In the rest of this paper we restrict ourselves to estimation around the true value $\vec{\varphi} = [0,0,0]$ to simplify algebraic expressions.
This is a relevant setting in precision magnetometry, where small deviations from a known reference field are measured.
Furthermore, this is usually the optimal point in parameter space and can be achieved by interspersing the evolution with adaptive control unitaries~\cite{Yuan2016b,Chen2019}, although a formal proof is lacking for the HCRB. 

\subsection{Evaluation of the HCRB}
We start with a real valued two-qubit state,
\begin{equation}
\ket{\psi_0} =
\begin{bmatrix}
r_1\\
r_2\\
r_3\\
r_4
\end{bmatrix},
\label{eq:psi0}
\end{equation}
such that $r_i \in \mathbb{R}$ and $\sum_i r_i^2 = 1$.
Whilst the issue of ``how real'' is quantum metrology has yet to be answered rigorously~\cite{Aaronson}, we have strong numerical evidence that the optimal state is real for two-qubit 3D magnetometry.

For this problem the vectors~\eqref{eq:horlift}, stacked as columns of a matrix, are
\begin{align}
& \left( \ket{l_1} \, \ket{l_2} \, \ket{l_3} \right) \\
& = \left(
\begin{array}{ccc}
2 (2 r_1 \rofp-1) (\rttp) & 2 \I \rttp & 4 r_1 \left(\rofp \rofm-1\right)  \\
2 \rofp (2 r_2 \rttp-1) & -2 \I \rofm & 4 r_2 \rofm \rofp \\
2 \rofp (2 r_3 \rttp-1) & -2 \I \rofm & 4 r_3 \rofm \rofp   \\
2 (2 r_4 \rofp-1) \rttp & -2 \I \rttp & 4 r_4 \left(\rofp \rofm+1\right)  \\
\end{array}
\right),\nonumber
\end{align}
where we have introduced
\begin{equation}
r_{1,4}^\pm = r_1 \pm r_4, ~~~~ r_{2,3}^\pm = r_2 \pm r_3.
\end{equation}
As $\dim \text{span}_\mathbb{R}\{\ket{l_i} \} = 3$ and $\dim \text{span}_\mathbb{C}\{\ket{l_i} \} = 2$,

\begin{equation}
\begin{split}
\spn_\mathbb{C}\{\ket{l_i} \} &= \spn_\mathbb{C} \{\ket{l_1},\ket{l_2}\}\\
&\cong \spn_\mathbb{R}\{\ket{l_1}, \ket{l_2},i\ket{l_1}, i\ket{l_2} \}\\
&\cong \spn_\mathbb{R}\{\ket{l_1}, \ket{l_2},\ket{l_3}, \ket{v} \},
\end{split}
\end{equation}
where 
$\ket{v}$ is a complex linear combination of  $\ket{l_1}$ and $\ket{l_2}$ satisfying
$\text{Re}\braket{l_i|v} = 0$.
Explicitly,
\begin{equation}
\ket{v} = -\frac{\I \left((\rofm)^2+(\rttp)^2\right)}{\rofp \rofm}\ket{l_1}-\ket{l_2}.
\end{equation}
By substituting the constraints~\eqref{eq:HCRBpureconst} we get to
\begin{equation}
\label{eq:xconst}
\ket{x_i} = \sum_j (J^{-1})_{ji}\ket{l_j}+ \alpha_i \ket{v},
\end{equation}
where now for pure states the QFI matrix~\eqref{eq:QFImat} is $J_{ij} = \Re \braket{l_i | l_j}$.
Crucially, we now have an unconstrained optimisation on the three real parameters $\alpha_i$, that is, $\CH = \min_{\vec{\alpha}} \Tr \Re Z [\vec{\alpha}] + \lVert \Im Z [\vec{\alpha}] \rVert_1$.
The same approach of explicitly substituting the constraints was initially applied to mixed states~\cite{Suzuki2018}, being instrumental in obtaining closed-form results for qubits~\cite{Suzuki2016a}.

This function is minimised by $\vec{\alpha} = [0,0,0]$ as can be explicitly checked by the vanishing of the gradient and the positive semidefiniteness of the Hessian (See details in Appendix~\ref{app:hessian}).
In others words, the optimal vectors $\ket{x_i}$ lie in the real subspace generated by the vectors $\ket{l_i}$ so we have
\begin{equation}
\label{eq:coherentbound}
\CH = C^S + \left\lVert \jsin D \jsin \right\rVert_1,
\end{equation}
where $D_{ij} =  \Im \Tr \left[ L_i L_j \rho_{\vec{\varphi}} \right] = \Im \braket{l_i | l_j }.$ 
More explicitly, the HCRB is
\begin{widetext}
\begin{align}
\label{equ:ch}
\CH = \frac{1}{8} \left(\frac{1} {  (\rofp) ^2 + (\rofm) ^2 -2 (\rofm \rofp)^2 + (\rttp)^2 (1 - 2 (\rofp)^2)} + \left(\frac{1}{\sqrt{(\rofm)^2+(\rttp)^2}}+\frac{1}{(\rofp)} \right)^2\right).
\end{align}
\end{widetext}

It is interesting to note that \eqref{eq:coherentbound} is in general an upper bound to the HCRB and only equal for any weight matrix if and only if the model is D-invariant~\cite{Suzuki2016a}.
For pure states and an even number of parameters, such models are called coherent as their tangent space has a symplectic structure~\cite{Fujiwara1999}.
For two-qubit 3D magnetometry we find \eqref{eq:coherentbound} to hold for any diagonal weight matrix but not for a general one (See Appendix~\ref{app:weight_maxt} for details).
This is consistent with the fact that it is not a coherent model.

Before moving onto the attainability of this bound we briefly comment on the role of entanglement in the initial state.
We find a one-to-one relationship between the entanglement of the input state and HCRB, with both separable and maximally entangled states leading to a singular model, meaning that they do not allow for the simultaneous estimation of all three parameters (See Appendix~\ref{app:role_of_ent} and Ref.~\cite{Baumgratz2015}).

\subsection{Attainability with projective measurements}
\label{subsec:projattain}

The HCRB for all pure state models are attainable with a POVM, without the need of measuring multiple copies collectively~\cite{Matsumoto2002}.
However, ancillas may be required to implement the optimal measurement.
The HCRB is attainable with a projective measurement if $\dim \mathcal{H}>2 p +1$~\cite[Chap. 20, Corollary 23]{Hayashi2005}. For two-qubit 3D magnetometry, $\dim \mathcal{H} = 4 < 2 p + 1 = 7$.
Thus, known results offer no suggestion of projective attainability of the HCRB for 3D magnetometry with two-qubit pure states.

We now show the projective attainability of the HCRB for 3D magnetometry with two-qubit real pure states.
To that end, recall that a projective measurement (and a locally unbiased estimator) with $V(\Pi,\tilde{\vec{\varphi}})_{ij} = F(\ket{\psi_{\vec{\varphi}}},\Pi)^{-1}_{ij} = \braket{x_i | x_j }$ exists if and only if $\Im \braket{x_i | x_j}  = 0$ for vectors also satisfying the local unbiasedness condition~\eqref{eq:HCRBpureconst}~\cite{Matsumoto2002}.
Since we have already found the HCRB, to prove its projective attainability we need to find a set $\{ \ket{x_i} \}$ satisfying~\eqref{eq:HCRBpureconst} as well as $\CH  = \Tr \Re Z[X] \quad \text{and} \quad \Im Z[X] = 0.$ Unlike the evaluation of the HCRB, it is not possible to solve this system by restricting to $ \ket{x_i} \in \spn_{\C} \{ \ket{l_1},\ket{l_2} \}$.
Thus, we have 24 real parameters (coming from the 3 complex 4-dimensional vectors) with 6 linear constrains, 3 bi-linear constraints and 1 quadratic constraint.
Indeed we are able to satisfy all these constraints; a family of $X$ operators that attains the HCRB for 3D magnetometry with two-qubit real pure states can be found in Appendix~\ref{app:pure}.

\section{HCRB for noisy two-qubit states}
\label{sec:noisy}

In the real world, quantum-limited 3D magnetometry will be noisy, making an initially pure probe mixed. 
Attaining the HCRB for mixed states is intimately tied to collective POVMs across multiple copies of the states~\cite{Demkowicz-Dobrzanski2020}.
Furthermore, the optimal asymptotic collective measurement identified by the theory of quantum local asymptotic normality is not projective~\cite{Demkowicz-Dobrzanski2020}.
This implies the need for noiseless ancillas which is theoretically typical~\cite{Demkowicz-Dobrzanski2014,Huang2016a,Sekatski2016,Zhou2017,Wang2018b,Sbroscia2018,Gorecki2019} but practically impossible.

To identify practically feasible pathways for attaining the HCRB for noisy two-qubit 3D magnetometry, we evaluate the scalar and channel bounds depicted in Fig.~\ref{fig:crbs-cd} for projective measurements and a few copies as opposed to POVMs and infinitely many copies.
This motivates the definition of the \emph{projectively attainable} $k$-copy bound as (cf. \eqref{eq:kcopyCRB})
\begin{equation} 
\label{eq:projkcopyCRB}
C^{(k)}_*\of{\rho_{\vec{\varphi}}}= \min_{\text{proj. } \Pi^{(k)}} k \, \CC \of{\rho_{\vec{\varphi}}^{\otimes k},\Pi^{(k)} }.
\end{equation}
Practical considerations have thus inspired mathematical quantities which deserve attention.
Indeed, all quantities defined in Fig.~\ref{fig:crbs-cd} via minimisations over POVMs now admit a projective equivalent denoted by a $*$ in the subscript.
Evidently, such projective channel bounds are greater or equal than the unrestricted versions, i.e. $C^{(k)}_* \geq C^{(k)}$.
Our central objective in this section is to study the approach of $C^{(k)}_*\of{\rho_{\vec{\varphi}}}$ towards $\CH \of{ \rho_{\vec{\varphi}} }$ for small $k$.

Our small $k$ investigations will rely on numerical optimisations and, as with all non-convex optimisation problems, it is in general very hard to guarantee the global optimality of the solution.
As mitigation, we use two independent parametrisations and optimisation methods. 

For 3D magnetometry, we focus on independent and identical dephasing (in the $z$ direction, without loss of generality) as our dominant noise process and parametrise it for a single qubit as~\cite{nielsen2010quantum} 
\begin{equation}
\Lambda_\gamma[\rho] = \sum_i E_i \rho E_i^\dagger,
E_0 = \begin{bmatrix}1 & 0 \\ 0 & \sqrt{1-\gamma} \end{bmatrix},
 E_1 = \begin{bmatrix}0 & 0 \\ 0 & \sqrt{\gamma} \end{bmatrix},
 \label{eq:deph}
\end{equation}
for $ \gamma \in [0,1].$ $\gamma =0$ denotes no noise while $\gamma=1$ denotes complete dephasing.
Under this noise, an $N$-qubit initial state $ \rho_0$ evolves into
\begin{equation}
\rho_{\vec{\varphi}, \gamma} = \mathcal{E}_{\vec{\varphi},\gamma}^{\otimes N} [ \rho_0 ] = \Lambda_\gamma^{\otimes N} \left[ U(\vec{\varphi})^{\otimes N }  \rho_0 U(\vec{\varphi})^{\otimes N \dagger} \right],
\label{eq:rhoN}
\end{equation}
where we denote the combined single-qubit channel for noisy 3D magnetometry as $\mathcal{E}_{\vec{\varphi},\gamma}$.

In the following, we use numerical optimisation to explore arbitrary levels of noise.
Based on these numerics, we conjecture that two copies are enough to attain the HCRB for high noise and that there is almost no advantage to adding a small number of copies at low noise.

\subsection{\label{subsec:numerical_results}Results from numerical optimisation}

To understand the attainability of the HCRB for 3D magnetometry at intermediate values of noise, we compute the channel bounds $\bar{C}^{\textrm{H}}\of{\mathcal{E}_{\vec{\varphi},\gamma}^{\otimes{2}} }$ and $\tilde{C}_*^{(k)}\of{\mathcal{E}_{\vec{\varphi},\gamma}^{\otimes{2}} }$~\footnote{
Note that we consider $k$ copies of two qubit systems.
Thus, since the encoding acts identically on single qubits, we have to use the two-qubit channel $\mathcal{E}_{\vec{\varphi},\gamma}^{\otimes{2}}$ in the arguments of the bounds in Fig.~\ref{fig:crbs-cd}.
Seeking to avoid the fate of Von Neumann's onions~\cite{VonNeumann}, we resist the temptation of introducing even more notation.}.
See Fig.~\ref{fig:crbs-cd}.
Both these quantities require finding the HCRB-optimal single-copy state and the optimal $k$-copy projective measurements---the latter not optimised over initial states, but evaluated for the HCRB-optimal state.
In the absence of any analytical techniques, we use numerical optimisation algorithms to find them for each considered value of $\gamma$ and limit ourselves to $k=2,3$.

We begin by paramatrizing the set of projective measurements and initial states for the optimisation.
Since we work with a multi-qubit system, we utilise a Bloch-vector parametrisation of a Hamiltonian. 
That is, the Hamiltonian is formed from a linear combination of $\text{SU}(4 k )$ generators obtained by tensoring $2 k$ $\text{SU}(2)$ generators, namely, $\lambda(k) =  \{\mathbb{I},\sigma_x,\sigma_y,\sigma_z   \}^{\otimes 2 k} \setminus \mathbb{I}^{\otimes 2 k} .$ The corresponding unitary is

\begin{equation}
\label{eq:unitary}
U(k, \Tilde{\vec{\alpha}}) = \exp\left( \I \sum_{i=1}^{16^{k}-1} \Tilde{\alpha}_{i} \lambda(k)_i  \right).
\end{equation} 
An initial single-copy pure state is then given by $U(1, \Tilde{\vec{\alpha}}_\psi)\ket{0}$, starting from a fiducial two-qubit state $\ket{0} \in \mathbbm{C}^4$; we then optimise over the coefficients $\Tilde{\vec{\alpha}}_\psi$ to find the optimal state $\ket{\tilde{\psi}_0}$. 
This is the only state we seek to optimise here. 
Note that we do not restrict our numerical searches to real coefficients as in Eq.~\eqref{eq:psi0}.
A $k$-copy projective measurement is given as $\Pi = \{ U(k, \Tilde{\vec{\alpha}})| i \rangle \langle i | U(k, \Tilde{\vec{\alpha}})^\dag  \}_{i=1,...,4^{k}}$, with a fixed orthonormal basis $\{ \ket{i} \}$ of $\mathbbm{C}^{4k}$ (we choose the computational basis of the $2k$ qubits); again we need to optimise the coefficients $\Tilde{\vec{\alpha}}$.
Recall that first we find  $\ket{\tilde{\psi}_0}$ by minimizing $\CH\of{ \mathcal{E}_{\vec{\varphi},\gamma}^{\otimes{2}} , U(1, \Tilde{\vec{\alpha}}_\psi)\ket{0} }$. 
This gives the quantity $\bar{C}^{\textrm{H}}\of{\mathcal{E}_{\vec{\varphi},\gamma}^{\otimes{2}} }$. $\ket{\tilde{\psi}_0}$ is then used to find the optimal $k$-copy projective measurement by minimizing $k \, \CC\of{ \mathcal{E}_{\vec{\varphi},\gamma}^{ \otimes 2 k} , {\tilde{\ket{\psi_0}}}^{\otimes k} , \{ U ( k  , \Tilde{\vec{\alpha}}) | i \rangle \langle i | U ( k  , \Tilde{\vec{\alpha}})
^\dag \} }$ as the objective function.
This leads to $\tilde{C}^{(k)}\of{\mathcal{E}_{\vec{\varphi},\gamma}^{\otimes{2}} }$.
The details of the optimisation process are provided in Appendix~\ref{subsec:unitary}.

\begin{figure}[h!]
\includegraphics[width = \linewidth]{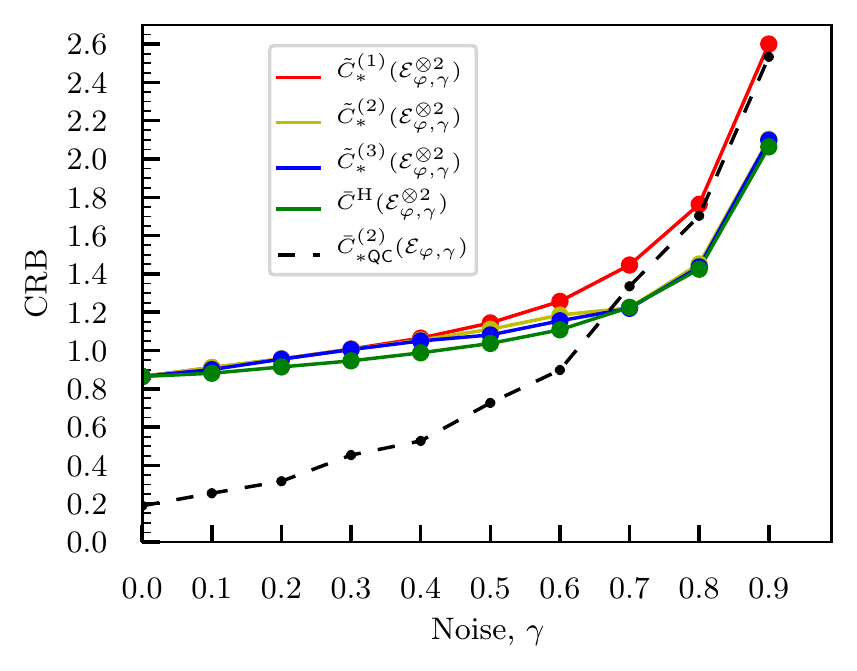}
\caption{
    Solid lines: channel HCRB $\bar{C}^{\textrm{H}}\of{\mathcal{E}_{\vec{\varphi},\gamma}^{\otimes{2}}  }$ (optimised over initial state) and $k$-copy projective bounds $\tilde{C}_*^{(k)}\of{\mathcal{E}_{\vec{\varphi},\gamma}^{\otimes{2}}  }$ (optimised over projective measurements on multiple copies and computed for the HCRB-optimal initial state, see Fig.~\ref{fig:crbs-cd}).  
    These solid lines correspond to the CQ strategy (e.g. left diagram in Fig.~\ref{fig:CQvsQCcirc} for $k=2$).
    Dashed line: CRB $\bar{C}^{(2)}_{*\mathsf{QC}}\of{\mathcal{E}_{\vec{\varphi},\gamma} }$ for the QC strategy for $k=2$ (right diagram in Fig.~\ref{fig:CQvsQCcirc}, optimised over 4-qubit initial states and over 2-qubit projective measurements).
    }
	\label{FIG:opt_state_and_measure}
\end{figure}

The results from the numerical optimisation are shown as the solid lines in Fig.~\ref{FIG:opt_state_and_measure}. 
The red, blue and light green solid lines show the quantity $\tilde{C}_*^{(k)}\of{\mathcal{E}_{\vec{\varphi},\gamma}^{\otimes{2}}  }$ for $k=1,2,3$ respectively.
These results are compared against the channel HCRB $\bar{C}^{\textrm{H}}\of{\mathcal{E}_{\vec{\varphi},\gamma}^{\otimes{2}}}$, computed by numerically maximizing the HCRB (evaluated by solving a semi-definite program~\cite{Albarelli2019}), and presented as a green line.
We see that adding a second and third copy of the state gives almost no advantage in the low-noise regime, while at the high-noise end we see that two-copies of the state effectively attains the HCRB and adding the third copy does not give an additional advantage. 
The intermediate-noise regime shows distinction between the CRBs of the one-copy, two-copies, and three-copies measurement schemes, although respective advantages are small.

\subsection{High and low noise: Analytical conjectures}
\label{subsec:highnoise_anal}
Spurred by our numerical results we conjecture that for $\gamma \approx 1$,
\begin{equation}
\tilde{C}_*^{(2)}\of{\mathcal{E}_{\vec{\varphi},\gamma}^{\otimes{2}} }\approx \bar{C}^{\textrm{H}}(\mathcal{E}_{\vec{\varphi},\gamma}^{\otimes{2}}),
\label{eq:highnoiseConj}
\end{equation}
that is for large $\gamma$, the HCRB is attainable with a projective measurement on two copies of the state.
This conjecture implies $\tilde{C}_*^{(2)} ( \mathcal{E}_{\vec{\varphi},\gamma}^{\otimes 2}) \approx \bar{C}_*^{(2)} ( \mathcal{E}_{\vec{\varphi},\gamma}^{\otimes 2}) \approx \bar{C}^{(2)} ( \mathcal{E}_{\vec{\varphi},\gamma}^{\otimes 2})$.
In words, we also conjecture that in the high-noise limit and for $k \geq 2$ copies non-projective measurements are not necessary to attain the channel bound $\bar{C}^{(k)}$.

Further, for $\gamma \approx 0$,
\begin{equation}
    \tilde{C}_*^{(1)}\of{\mathcal{E}_{\vec{\varphi},\gamma}^{\otimes{2}} }\approx \tilde{C}_*^{(2)}\of{\mathcal{E}_{\vec{\varphi},\gamma}^{\otimes{2}} },
\end{equation}
meaning that for small $\gamma$ adding one additional copy gives no substantial advantage over the one copy case. Based on our numerical observations, we further conjecture that
\begin{equation}
\tilde{C}_*^{(2)}\of{\mathcal{E}_{\vec{\varphi},\gamma}^{\otimes{2}} } > \bar{C}^{\textrm{H}}(\mathcal{E}_{\vec{\varphi},\gamma}^{\otimes{2}}),
\end{equation}
that is, there still remains a gap between $\tilde{C}_*^{(2)}$ and $\bar{C}^{\textrm{H}}$ for low noise unlike Eq.~\eqref{eq:highnoiseConj} for high noise.
Some intuition may be obtained by noting that for $\gamma \approx 0$, the two-copy state $\rho_{\vec{\varphi}, \gamma}^{\otimes 2}$ is very close to being pure for which the HCRB is attainable with a projective measurement.
Hence one additional copy provides little benefit.
However, for increasing $k$ the mixedness of $\rho_{\vec{\varphi}, \gamma}^{\otimes k}$ increases monotonically, explaining the gap between  $\tilde{C}_*^{(2)}$ and $\bar{C}^{\textrm{H}}$.

\begin{figure}[h!]
\[
\Qcircuit @C=1em @R=0.5em {
	\lstick{}&\gate{\mathcal{E}_{\vec{\varphi},\gamma}}&\multimeasureD{3}{}
	\\
	\lstick{}&\gate{\mathcal{E}_{\vec{\varphi},\gamma}}&\ghost{}
	\inputgroupv{1}{2}{0.8em}{1em}{\ket{\psi_0}}
	\\
	\lstick{}&\gate{\mathcal{E}_{\vec{\varphi},\gamma}}&\ghost{}
	\\
	\lstick{}&\gate{\mathcal{E}_{\vec{\varphi},\gamma}}&\ghost{}
	\inputgroupv{3}{4}{0.8em}{1em}{\ket{\psi_0}}
}
\qquad \qquad
\Qcircuit @C=1em @R=0.5em {
	\lstick{}&\gate{\mathcal{E}_{\vec{\varphi},\gamma}}&\multimeasureD{1}{}
	\\
	\lstick{}&\gate{\mathcal{E}_{\vec{\varphi},\gamma}}&\ghost{}
	\\
	\lstick{}&\gate{\mathcal{E}_{\vec{\varphi},\gamma}}&\multimeasureD{1}{}
	\\
	\lstick{}&\gate{\mathcal{E}_{\vec{\varphi},\gamma}}&\ghost{}
	\inputgroupv{1}{4}{0.8em}{3.0em}{\ket{\psi_0}}
}
\]

\caption{Classical-quantum (CQ) (left) vs quantum-classical (QC) (right) schemes for $k=2$ two-qubit systems; each wire represent a single qubit system.
The channel $\mathcal{E}_{\vec{\varphi},\gamma}=  \Lambda_\gamma \circ \mathcal{U}_{\vec{\varphi}}$ represents the noisy encoding and acts independently on each qubit; $\Lambda_\gamma$ is given by Eq.~\eqref{eq:deph}, $\mathcal{U}_{\vec{\varphi}}$ is the unitary encoding of the parameters $\vec{\varphi},$ 
and $\ket{\psi_0}$ denotes the initial states.}
\label{fig:CQvsQCcirc}

\end{figure}
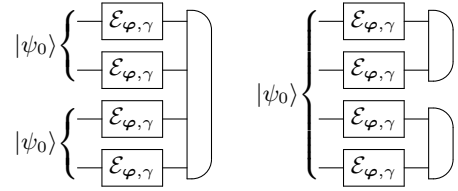

\subsection{CQ versus QC strategies}

The $k$-copy collective measurements that we have considered so far are operationally implemented by an entangling unitary operation on $2k$ qubits.
A natural question is whether it is better to apply this global unitary at the preparation stage before recording the parameter or at the measurement stage after recording the parameter.
In other words the strategies considered so far to study how the HCRB is approached are classical-quantum (CQ) ones.
We now explore quantum-classical (QC) strategies, in which the measurements are local on the two-qubit systems, but the initial states contains quantum correlations across all $2k$ qubit subsystems.
In particular, we explore numerically the first non-trivial case $k=2$, schematically depicted in Fig.~\ref{fig:CQvsQCcirc}.
Ours are the first investigations on the best place to use quantum correlations in noisy multiparameter quantum estimation.

The quantity capturing the performance of QC strategies is not depicted in Fig.~\ref{fig:crbs-cd} and reads as
\begin{equation}
\label{eq:QCbound}
\bar{C}^{(2)}_{\mathsf{QC}}\of{\mathcal{E}_{\vec{\varphi},\gamma} } = \min_{ \ket{\psi_0} , \Pi} 2 \, C\of{ \mathcal{E}_{\vec{\varphi},\gamma}^{\otimes 4} , \ket{\psi_0} , \Pi^{\otimes 2}},
\end{equation}
where now, since $k=2$ we have $\ket{\psi_0} \in \mathbbm{C}^{8}$ and $\Pi $ is a 2-qubit projective measurement over $\mathbbm{C}^4$.
This should be compared against the quantity $\tilde{C}^{(2)}$ corresponding to a CQ scheme where a 4-qubit unitary is employed after the parameter encoding.
Ideally $\bar{C}^{(2)}_{\mathsf{QC}}\of{\mathcal{E}_{\vec{\varphi},\gamma}}$ should also be compared against $\bar{C}^{(2)}$, however after a preliminary numerical check we found effectively no difference to $\Tilde{C}^{(2)}$, as one might intuitively expect, and therefore only considered the latter quantity, being computationally less demanding to evaluate.

The quantity~\eqref{eq:QCbound} is plotted with a dashed line in Fig.~\ref{FIG:opt_state_and_measure}.
For low and medium noise the QC strategy outperforms the CQ strategy. This is to be expected, at least for low noise, from previous results in the purely unitary case~\cite{Baumgratz2015}.
However, at higher values of $\gamma$ we see that the increasing impact of noise on the collective state brings down the performance of the QC strategy almost to the two-qubit single-copy case.
In this regime the advantage from a collective measurement can be appreciated revealing a unique aspect of quantum multiparameter estimation.

\begin{figure*}[th!]
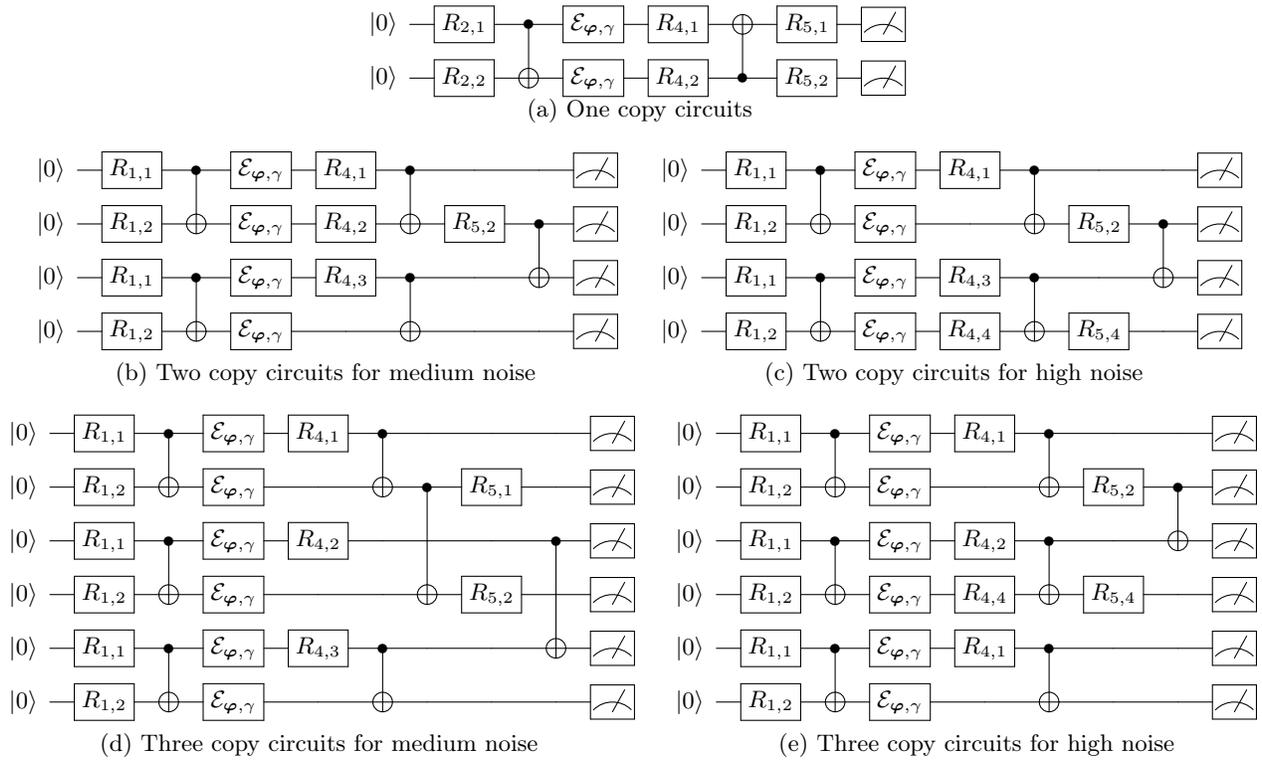

    \begin{minipage}{\textwidth}
          \centering
          \scalebox{1}{\include{figures/implemnt_circuits/one_copy} }\\
          (a) One copy circuits
          ~\vspace{10pt}
    \end{minipage}
     \begin{minipage}{.45\textwidth}
          \scalebox{1}{
          \include{figures/implemnt_circuits/two_copy_medium}  }\\~\\
          (b) Two copy circuits for medium noise
          ~\vspace{10pt}
    \end{minipage}
    \hspace{2pt}
     \begin{minipage}{.45\textwidth}
          \scalebox{1}{\include{figures/implemnt_circuits/two_copy_high}  }\\~\\
          (c) Two copy circuits for high noise
          ~\vspace{10pt}
    \end{minipage}
    ~\vspace{10pt}
         \begin{minipage}{.45\textwidth}
         \scalebox{1}{ \include{figures/implemnt_circuits/three_copy_medium}  }\\~\\
          (d) Three copy circuits for medium noise
    \end{minipage}
    ~\hspace{10pt}
     \begin{minipage}{.45\textwidth}
          \scalebox{1}{\include{figures/implemnt_circuits/three_copy_high_noise}  }\\~\\
          (e) Three copy circuits for high noise
    \end{minipage}
	\caption{
	Optimal quantum circuits for 3D magnetometry across noise strengths and number of copies.
	The final single-qubit measurement represented by the meter box is a standard projection on $\sigma_z$ eigenstates.
	For zero noise (a), the HCRB is attainable projectively on one copy as shown in Sec. \ref{subsec:projattain}.
	The circuit shown in (a), is of this form for all noise values with one copy.
	In (b), (c), the CNOT between the two copies for both medium and high noise shows the usefulness of the second copy.
	This provides an independent route to the conclusion of Fig.~\ref{FIG:opt_state_and_measure}. 
	In (d), the CNOTs between the three copies at medium noise show that all three need to be entangled to approach the HCRB.
	In (e), the lack of a CNOT to the third copy shows that it is unnecessary in attaining the HCRB, an independent reinforcement of another conclusion of Fig.~\ref{FIG:opt_state_and_measure}.
	Medium noise optimisation is performed for $\gamma = 0.5$ and high noise for $\gamma = 0.8$.
	}
	\label{fig:circ}
\end{figure*}

\section{Shallow circuits approaching HCRB}
\label{sec:circ}

In the spirit of moving from issues of principle to those of practice, we now present experimental pathways to attaining the lower bounds computed in~Sec.~\ref{subsec:numerical_results} using noisy intermediate-scale quantum devices~\cite{Nash2020}.
To that end, we use programmable quantum circuits as representation of the unitary operators.
There have been successful applications of machine learning and numerical optimisation for quantum estimation using various types of parametrised circuits~\cite{ODriscoll2019,Yang2020a,VanStraaten2020,Meyer2020}.
We base our circuits on the programmable universal quantum circuit proposed by Sousa and Ramos~\cite{Benicio}.
See Appendix~\ref{subsec:Qcircuit} for details.

To optimise the circuit, we split the parameterisation of our circuits into a discrete set and a continuous set.
The discrete part is captured by two bit strings representing the single-qubit and two-qubit gates being present or absent.
The continuous part parameterises the three Pauli gates forming a single-qubit rotation. 
The numerical optimisation is performed by combining a genetic-inspired algorithm and a gradient-descent algorithm.
The generic algorithm was originally devised for discrete optimisation~\cite{Wu1995}, thus suited for the discrete part of the parameter space.
The gradient-descent performs a local search on continuous search space.
Details of both algorithms are given in Appendix~\ref{subsec:Qcircuit}.

Our results in Fig.~\ref{fig:circ} show that whether one, two or three copies are used, the circuits are relatively shallow in depth. This is also true for alternative optimisation strategies considered in Appendix~\ref{app:Optimised quantum circuits}.
The state generation circuits, for example, only consists a single CNOT gate for each copy of the state.

\section{Discussion and open problems}\label{sec:concl}

We have undertaken the first analysis of attainable multiparameter bounds for quantum-limited 3D magnetometry---the estimation of three phase parameters.
We have provided the first analytical expression for the HCRB of a three parameter estimation problem using two-qubit states.
We also studied numerically the attainment of the HCRB for noisy two-qubit 3D magnetometry using collective projective measurements on a few copies of the state.
Finally, we provided shallow quantum circuits for 3D magnetometry approaching the fundamental quantum limit set by the HCRB using up to three copies of two-qubit states.

This paper is a first step towards bringing quantum-limited 3D magnetometry, and multiparameter quantum estimation more generally, to fruition.
It opens many interesting questions for future research some of which we highlight below.
	\begin{enumerate} 
		\item Attainability of the HCRB by projective measurements for pure states and with few copies for mixed states. 
		
		\item Asymptotic attainability of the HCRB by projective measurements. This translates to investigating the closeness of
		\begin{equation}
\label{eq:CprojHlim}
C_*^{(\infty)} ( \rho_{\vec{\varphi}}) = \lim_{k \rightarrow \infty}  C^{(k)}_*\of{\rho_{\vec{\varphi}}}
\end{equation}
		 to $\CH \of{ \rho_{\vec{\varphi}} }.$ 
		\item A thorough comparison of CQ and QC strategies, with proper accounting of the cost of ancillas in such comparisons even when they enter only with non-projective measurements (more accurately, with POVMs that cannot be projectively simulated~\cite{Oszmaniec2016b}).
		\item A rigorous analysis into the usefulness of entangling untiaries in the most general QQ case.
		\item Further analysis into relationships presented Figure~\ref{fig:crbs-cd} including the extensions into the QQ and QC regimes. 

	\end{enumerate}
We close with the belief that addressing these open questions call new methods to assess non-asymptotic bounds in multiparameter quantum estimation theory.

\begin{acknowledgments}
We thank E.~Bisketzi, D.~Branford, R.~Demkowicz-Dobrzański, A.~Fujiwara and W.~Górecki  for useful discussions.
This work has been supported by the UK EPSRC (EP/K04057X/2), the UK National Quantum Technologies Programme (EP/M01326X/1, EP/M013243/1) and Centre for Doctoral Training in Diamond Science and Technology (EP/L015315/1).
FA acknowledges financial support from the National Science Center (Poland) grant No. 2016/22/E/ST2/00559.
\end{acknowledgments}

\bibliography{bibliography}

\clearpage
\onecolumngrid

\appendix

\section{Details on the HCRB for noiseless systems}

\subsection{\label{app:mag_details} 3D Magnetometry details}

In this work we are interested in the estimation of the 3 components of a magnetic field.
For two qubits, the Hamiltonian is given by
\begin{equation}
H(\vec{\varphi}) = h(\vec{\varphi}) \otimes \mathbb{I} + \mathbb{I} \otimes h(\vec{\varphi}) = \varphi_1(\sigma_x\otimes \mathbb{I}+ \mathbb{I}\otimes \sigma_x)+\varphi_2(\sigma_y\otimes \mathbb{I}+ \mathbb{I}\otimes \sigma_y)+\varphi_3(\sigma_z\otimes \mathbb{I}+ \mathbb{I}\otimes \sigma_z),
\end{equation}
where we introduced the standard Pauli matrices
\begin{equation}
\sigma_z = \begin{bmatrix}
1 & 0 \\
0 & -1
\end{bmatrix} \quad
\sigma_x = \begin{bmatrix}
0 & 1 \\
1 & 0
\end{bmatrix} \quad
\sigma_y = \begin{bmatrix}
0 & \I \\
-\I & 0
\end{bmatrix}.
\end{equation}
With subsequent evolution,
\begin{equation}
\ket{\psi_{\vec{\varphi}}} = e^{- \I  H(\vec{\varphi}) }\ket{\psi_0},
\end{equation}
where $\ket{\psi_0}$ is some pure input state. Importantly we also need the derivatives of the evolved state,
\begin{align}
\frac{\partial \ket{\psi_{\vec{\varphi}}}}{\partial\varphi_i} &= \frac{\partial}{\partial\varphi_i} e^{-\I H(\vec{\varphi}) }\ket{\psi_0} = -\I e^{- \I H(\vec{\varphi})}A_i(\vec{\varphi})\ket{\psi_0},
\end{align}
with Hermitian operator,
\begin{equation}
A_i(\vec{\varphi}) = \int_0^1 e^{-\I \alpha H (\vec{\varphi})}H_ie^{ \I \alpha H(\vec{\varphi})} \text{d} \alpha.
\end{equation}
Where $ H_i$ is $\partial_{\varphi_i}  H(\vec{\varphi})$.
In this work we will in the the limit $\varphi_i \rightarrow 0 \, \forall i$, 
In this regime we can see that,
\begin{equation}
\lim_{\varphi_i \rightarrow 0}\frac{\partial}{\partial\varphi_i} e^{- \I H(\vec{\varphi})}\ket{\psi_0} = - \I H_i \ket{\psi_0}.
\end{equation}

\subsection{\label{app_asym_clas} Asymptotic classicality}

Asymptotically classical models, adopting the terminology of~\cite{Suzuki2018} are those for which $C^S = C^H$, determined by the condition $D=0$, where $D_{ij} = \Im \Tr L_i L_j \rho_{\vec{\varphi}} $~\cite{Ragy2016}.
For the problem we are considering, a sufficient condition for asymptotic classicality is to satisfy~\cite{Baumgratz2015},
\begin{equation}
\label{equ:asym_clas}
\Tr \rho^{[1]}a_ia_j = 0 \quad \forall i, j.
\end{equation}
For zero parameter $a_i \rightarrow \sigma_i \implies \Tr\rho^{[1]}a_ia_j = \varepsilon_{i,j,k}\alpha_k $, where $\rho^{[1]} = (\mathbb{I}+\sum \alpha_i \sigma_i)/2$.
The only way this is zero $\forall i,j$ is for $\alpha_i$ to be zero $\forall i \implies \rho^{[1]} = \mathbb{I}/2$.
There are only two classes of two-qubit pure states which have this 1-body reduced density matrix,
\begin{equation}
\ket{\psi_1} = \frac{1}{\sqrt{2}}(\ket{00} \pm e^{- \I \phi}\ket{11}) \qquad \ket{\psi_2} = \frac{1}{\sqrt{2}}(\ket{10} \pm e^{-\I \phi}\ket{01}),
\end{equation}
for some arbitrary phase, $e^{-\I \phi}$.
These two states are unable to estimate all three parameters, since they give rise to singular statistical models, for which the QFI matrix is not invertible.

\subsection{Convexity of unconstrained optimisation problem}\label{app:hessian}
We begin with the form of the $\ket{x_i}$ operators,
\begin{equation}
\label{app_eq:xconst}
\ket{x_i} = \sum_j (J^{-1})_{ji}\ket{l_j}+ \alpha_i \ket{v},
\end{equation}
where $\alpha_i$ is the vector of 3 real parameters that we minimise over. Restated, $\CH = \min_\alpha \Tr \Re Z [\vec{\alpha}] + \lVert \Im Z [\vec{\alpha}] \rVert_1$. 
Firstly, denoting $ \ket{\tilde{l_i}} = \sum_j (J^{-1})_{ji}\ket{l_j} $,
\begin{equation}
    \Tr \Re Z [\alpha] = \sum_i \alpha_i^2 \braket{v|v} + \alpha_i (2 \Re \braket{\tilde{l_i}|v}) + \braket{\tilde{l}_i|\tilde{l}_i},
\end{equation}
which is quadratic, hence convex in $\alpha_i$. 

Secondly, 
\begin{equation}
    \lVert \Im Z [\alpha] \rVert_1  = \frac{1}{2} \sqrt{\braket{x_1|x_2}^2 +\braket{x_1|x_3}^2 +\braket{x_3|x_2}^2 } .
\end{equation}
In order to show this is convex wrt $\alpha_i$ we show that the Hessian of this function is positive semi-definite (PSD), namely,
\begin{align}
    y\begin{bmatrix}
    \frac{\partial^2 \lVert \Im Z [\alpha] \rVert_1}{\partial\alpha_1^2} & \frac{\partial^2 \lVert \Im Z [\alpha] \rVert_1}{\partial\alpha_1 \partial \alpha_2} & \frac{\partial^2 \lVert \Im Z [\alpha] \rVert_1}{\partial\alpha_1 \partial \alpha_3}\\
    \frac{\partial^2 \lVert \Im Z [\alpha] \rVert_1}{\partial\alpha_2 \partial \alpha_1} & \frac{\partial^2 \lVert \Im Z [\alpha] \rVert_1}{\partial\alpha_1 ^2} & \frac{\partial^2 \lVert \Im Z [\alpha] \rVert_1}{\partial\alpha_2\partial \alpha_3}\\
    \frac{\partial^2 \lVert \Im Z [\alpha] \rVert_1}{\partial\alpha_3 \partial \alpha_1} & \frac{\partial^2 \lVert \Im Z [\alpha] \rVert_1}{\partial\alpha_3\partial\alpha_2 } & \frac{\partial^2 \lVert \Im Z [\alpha] \rVert_1}{\partial \alpha_3^2}\\
    \end{bmatrix}y^\dagger \geq 0,
\end{align}
for an arbitrary real-valued vector, $y = (y_1,y_2,y_3)$. A necessary and sufficient condition for the above to be true is all eigenvalues of the Hessian are non-negative. Indeed we find the eigenvalues $\lambda_i$ to be
\begin{equation}
    \lambda_1 = 0,
   ~~~ \lambda_2 =\frac{(\rttp)^2}{(4\rofm\rofp)^2},
   ~~~ \lambda_3 =\frac{64 (\rttp)^4 \delta ((\alpha_3 \rttp + \alpha_1 \rofm)^2+\alpha_2^2 \delta ) +(\rofm)^2}{ (2\rofm \rofp)^4},
\end{equation}
where $\delta = 1- 2( r_1 r_4 - r_2 r_3  )$.
Since $2|r_1 r_4 - r_2 r_3 |$ is the concurrence of a pure two-qubit state~\cite{concur}, which takes value in the range $\left[0,1\right]$, we see that $\delta \geq 0$.
Given the eigenvalues of the Hessian are non-negative, $\lVert \Im Z [\alpha] \rVert_1$ is a convex function of the parameters $\alpha_i$.
As $\Tr \Re Z [\alpha]$ is also convex with respect to $\alpha_i$, and the sum of two convex functions is convex, $\Tr \Re Z [\alpha] + \lVert \Im Z [\alpha] \rVert_1$ is convex with respect to $\alpha_i$ and the global minimum is found for $\alpha_i = 0 \; \forall \; i$ since the gradient vanishes there.

\subsection{\label{app:weight_maxt} HCRB for diagonal weight matrices } 
 
We introduce the weight matrix $W$, a $p \times p$ real-valued positive matrix.
The corresponding HCRB is obtained by generalizing Eq.~\eqref{eq:HCRB} to
\begin{equation}
\Tr W  V_{\vec{\varphi}} (\Pi, \Tilde{\vec{\varphi}}) \geq \CH \left(\proj{\psi}\right)  = \min_{X } \Tr \Re W Z[X] + \left\lVert \sqrt{W} \Im Z[X] \sqrt{W} \right\rVert_1,
\end{equation}
with the same constraints~\eqref{eq:HCRBconst} and the same definition~\eqref{eq:HCRBZmat}.
We identify two natural classes of weight matrices.
The first is a diagonal $W = \operatorname{diag}(w_1, w_2, w_3),$ $w_i \geq 0,$ $w_i \in \mathbb{R}$.
This corresponds to modifying the proportion to which we estimate each parameter.
The second class is the general non-diagonal matrix $W \ge 0$, which corresponds to taking linear combinations of the parameters.
For 3D magnetometry with diagonal weight matrices,
\begin{equation}
\label{eq:coherentboundW}
\CH \left( \proj{\psi}  \right)  = \Tr W J^{S-1} + \left\lVert \sqrt{W} J^{S-1} D J^{S-1} \sqrt{W} \right\rVert_1.
\end{equation}
In particular, 
\begin{align}
\text{Abs}\Im \sqrt{W} \jsin D\jsin \sqrt{W}  & = \frac{1}{4} \sqrt{\frac{w_1 w_2 (\rofm)^2+w_2 w_3 (\rttp)^2}{(\rofp)^2
		\left((\rofm)^2+(\rttp)^2\right)^2}}\\
8\Re W Z[X]  &=  \frac{\alpha}{\beta}.
\end{align}
Where
\begin{align}
    \alpha & = w_2\frac{1}{(\rofm)^2+(\rttp)^2}+ w_1 \left((\rofp)^2\left(r_1^2+r_4^2\right) \left(\left(r_1^2+r_4^2\right)
    \left(r_2^2+r_3^2-2\right)+\left(\rofm \rofp \right)^2+1\right)\right)\\
    &+w_3(((\rofp)^2 + (\rttp)^2 (1 - 2 (\rofp)^2 ))),\\
    \beta &= 2 (\rofp)^2 \left(r_1^2 \left(-r_1^2 \left(r_4^2+2\right)+r_1^4+4 r_4^2 (r_2 r_3+1)-r_4^4+1\right)+r_4^2 \left(r_4^2-1\right)^2\right)\\
    &+(\rttp)^2 \left(2 (\rofp)^2 \left(2
    (\rofp)^2+r_2^2+r_3^2-2\right)+1\right).
\end{align}
However, the form of Eq.~(\ref{eq:coherentboundW}) for the HCRB does not hold for a general weight matrix. 

\subsection{\label{app:role_of_ent} Role of entanglement}
\begin{center}
	\begin{figure}[t]
		\includegraphics[width=0.8\textwidth]{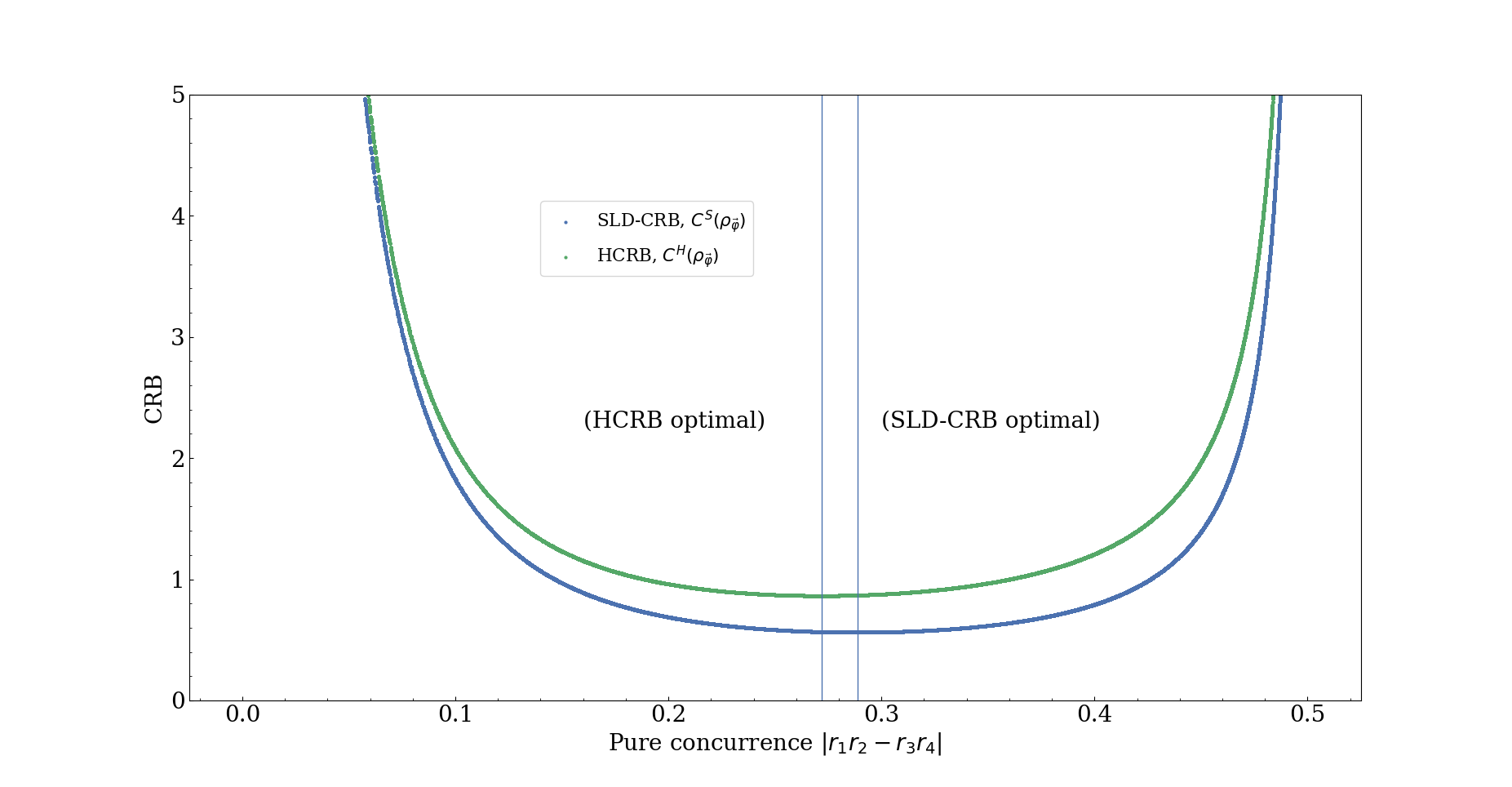}
		\caption{
		Parametric plot of (half the) concurrence~\cite{concur} $|r_1r_4-r_2r_3 |$ against the CRB, either the SLD one obtained from the QFI matrix or the HCRB.
		The vertical line to the left is the point of optimum HCRB and the one to the right is for the optimum SLD-CRB.}
		\label{fig:role of ent}
	\end{figure}
\end{center}

The role of entanglement as a resource in quantum metrology has been well-studied for single parameter problems~\cite{Apellaniz2015,Toth,Giovannetti1330}.
Thanks to the closed-form expression of the HCRB in \eqref{equ:ch}, QFI~\cite{Baumgratz2015}, and entanglement (as measured by the concurrence~\cite{concur}) for pure two qubit states of the form in Eq.~\eqref{eq:psi0}, we answer this question in full for pure-state two-qubit 3D magnetometry.
Fig.~\ref{fig:role of ent} shows a one-to-one relation between the entanglement and the HCRB for $\vec{\varphi}=0$.
The important point to note is that the HCRB and SLD-CRB obtained from the QFI are minimised for different amounts of entanglement.

\subsection{\label{app:pure}Attainability with projective measurements }

For pure states, the HCRB is given by (See \ref{sec:ps})
\begin{equation}
\CH \left(  \proj{\psi} \right) = \min_{X } \Re Z[X] + \text{Abs}\Im Z[X], ~~~~~~ Z[X] = XX^\dagger
\label{eqapp:HCRB}
\end{equation}
where $X= \left\lbrace\ket{x_1}, \ket{x_2},\ket{x_3} \right\rbrace$ is a matrix of vectors $X_i = \tilde{X}_i \ket{\psi} = \ket{x_i}.$
The operator  $\tilde{X}_i$ is an arbitrary Hermitian operator. The arbitrary complex vectors $\ket{x_i}$ satisfy the locally-unbiased condition as well as that their inner products are strictly real-valued, that is,
\begin{equation}
\label{app:equ:conditions}
\Re \braket{x_i|\partial_j \psi} = \delta_{i,j}, \quad \Im \braket{x_i|x_j} = 0. 
\end{equation}
We parametrise these vectors as
\begin{align}
X &= \left\lbrace\ket{x_1}, \ket{x_2},\ket{x_3} \right\rbrace \label{eqa:X} \\\
& = \begin{bmatrix}
x_{1,1}^\mathsf{r} + \I x_{1,1}^\mathsf{i} & x_{2,1}^\mathsf{r} + \I x_{2,1}^\mathsf{i} & x_{3,1}^\mathsf{r} + \I x_{3,1}^\mathsf{i} \\
x_{1,2}^\mathsf{r} + \I x_{1,2}^\mathsf{i} & x_{2,2}^\mathsf{r} + \I x_{2,2}^\mathsf{i} & x_{3,2}^\mathsf{r} + \I x_{3,2}^\mathsf{i} \\
x_{1,3}^\mathsf{r} + \I x_{1,3}^\mathsf{i} & x_{2,3}^\mathsf{r} + \I x_{2,3}^\mathsf{i} & x_{3,3}^\mathsf{r} + \I x_{3,3}^\mathsf{i} \\
x_{1,4}^\mathsf{r} + \I x_{1,4}^\mathsf{i} & x_{2,4}^\mathsf{r} + \I x_{2,4}^\mathsf{i} & x_{3,4}^\mathsf{r} + \I x_{3,4}^\mathsf{i}
\end{bmatrix},
\end{align}
which gives us 24 real parameters $x^{k}_{i,j}$ for $i=\{1,2,3\}, j = \{1,2,3,4\},k=\{\mathsf{r},\mathsf{i}\},$ nine linear constrains, three bi-linear constraints~\eqref{app:equ:conditions}, and one quadratic constraint $\CH \left(  \proj{\psi} \right)$~\eqref{eqapp:HCRB}, which leaves us with 11 free parameters. 

We begin with $\sum_{i =1}^3\braket{x_i|x_i}= \CH$, which gives,
\begin{equation}
x_{2,2}^\mathsf{r} = \sqrt{\CH - \braket{x_1|x_1} - \braket{x_3|x_3} - ((x_{2,1}^\mathsf{i})^2 + (x_{2,1}^\mathsf{r})^2 + (
	x_{2,2}^\mathsf{i})^2+ (x_{2,3}^\mathsf{i})^2 + (x_{2,3}^\mathsf{r})^2 + (x_{2,4}^\mathsf{i})^2 + (x_{2,4}^\mathsf{r})^2 )}
\end{equation}

Substituting the above in into $X$ (Eq.~\eqref{eqa:X}), satisfying $\braket{x_i|\partial_j \psi} = \delta_{i,j}$ gives 
\begin{align}
x_{1,3}^\mathsf{r} &= -(1/(r_{1,4}^+)) - ((r_{2,3}^+) x_{1,1}^\mathsf{r})/r_4 - x_{1,2}^\mathsf{r},\\
x_{1,4}^\mathsf{r} &= (r_1 x_{1,1}^\mathsf{r})/r_4,\\
x_{1,4}^\mathsf{i} &= x_{1,1}^\mathsf{i} - ((r_{1,4}^-) x_{1,2}^\mathsf{i})/(r_{2,3}^+) - ((r_{1,4}^-) x_{1,3}^\mathsf{i})/( r_{2,3}^+),\\
x_{2,3}^\mathsf{r} &= -(((r_{2,3}^+) x_{2,1}^\mathsf{r})/r_4) - x_{2,2}^\mathsf{r},\\
x_{2,4}^\mathsf{r} &= ( r_1 x_{2,1}^\mathsf{r})/r_4,\\
x_{2,4}^\mathsf{i} &= -(1/(r_{2,3}^+)) + x_{2,1}^\mathsf{i} - ((r_{1,4}^-) x_{2,2}^\mathsf{i})/(r_{2,3}^+) - ((r_{1,4}^-) x_{2,3}^\mathsf{i})/( r_{2,3}^+),\\
x_{3,3}^\mathsf{r} &= -((r_{2,3}^+)/(2 r_4 (r_{1,4}^+))) - ((r_{2,3}^+) x_{3,1}^\mathsf{r})/r_4 - x_{3,2}^\mathsf{r},\\
x_{3,4}^\mathsf{r} &= 1/(2 r_4) + (r_1 x_{3,1}^\mathsf{r})/r_4,\\
x_{3,4}^\mathsf{i} &= x_{3,1}^\mathsf{i} - ((r_{1,4}^-) x_{3,2}^\mathsf{i})/(r_{2,3}^+) - ((r_{1,4}^-) x_{3,3}^\mathsf{i})/(r_{2,3}^+).
\end{align}
Next satisfying $\Im \braket{x_j|x_j} = 0$ gives
\begin{align}
x_{1,1}^\mathsf{r} &= ((r_{2,3}^+) r_4 (x_{1,1}^\mathsf{i} x_{3,1}^\mathsf{r} + (((r_{2,3}^+) x_{1,1}^\mathsf{i} - (r_{1,4}^-) (x_{1,2}^\mathsf{i} +  x_{1,3}^\mathsf{i})) (1 + 2 r_1 x_{3,1}^\mathsf{r}))/(2 (r_{2,3}^+) r_4)\\ \nonumber & - ((r_{2,3}^+) x_{1,3}^\mathsf{i} (1 + 2 (r_{1,4}^+) x_{3,1}^\mathsf{r}))/(2 r_4 (r_{1,4}^+))- x_{1,2}^\mathsf{r} x_{3,2}^\mathsf{i} + x_{1,2}^\mathsf{i} x_{3,2}^\mathsf{r} - x_{1,3}^\mathsf{i} x_{3,2}^\mathsf{r}\\ \nonumber & + x_{3,3}^\mathsf{i}/( r_{1,4}^+) + x_{1,2}^\mathsf{r} x_{3,3}^\mathsf{i}))/(-r_1^2 (x_{3,2}^\mathsf{i} + x_{3,3}^\mathsf{i}) + (r_{2,3}^+) (r_4 x_{3,1}^\mathsf{i} - (r_{2,3}^+) x_{3,3}^\mathsf{i})\\\nonumber \nonumber & + r_1 ((r_{2,3}^+) x_{3,1}^\mathsf{i} + r_4 (x_{3,2}^\mathsf{i} + x_{3,3}^\mathsf{i}))),\\
x_{2,1}^\mathsf{i} &= 1/(r_{2,3}^+)\\
x_{1,1}^\mathsf{i} &= -((2 r_4)/(r_{1,4}^+)),
\end{align}
Finally, we set the remaining variables which are free to 1 or 0 to avoid singularities as
$x_{1,2}^\mathsf{i} =  x_{1,2}^\mathsf{r} = x_{1,3}^\mathsf{i} =  x_{2,1}^\mathsf{r} = x_{2,2}^\mathsf{i} = x_{2,3}^\mathsf{i} =  x_{3,1}^\mathsf{r}= x_{3,2}^\mathsf{r} = 0$ and $x_{3,3}^\mathsf{i} = x_{3,1}^\mathsf{i} =x_{3,2}^\mathsf{i} = 1.$
We set these to 0 or 1 for convenience, although one could spend more time choosing these parameters which give physically intuitive $X$ operators.

We have thus constructed a set of $X$ operators which gives rise to a real valued $Z\left[X\right]$ that also attains the HCRB. This implies that there exists projective measurement that attains the HCRB for two-qubit 3D magnetometry.

\section{\label{subsec:unitary} Numerical unitary optimisation}

In this Appendix, we detail the numerical methods used in Sec.~\ref{subsec:numerical_results}
to optimise the unitary operators that identify the optimal input states and projective measurements to numerically attain the HCRB and the $\CC$ for the medium and high noise.
As the number of parameters in a unitary operator in $\text{SU}(2)^{\otimes {2k}}$ grows exponentially with $k,$ we invoke the permutationally symmetric structure of the Hamiltonian to reduce size of the search space before implementing a particle swarm optimisation (PSO) algorithm to the remaining parameters.

\subsection{\label{subsubsec:unitary parameters}Parametrisation and permutational invariance}

The unitary as defined in Eq.~\eqref{eq:unitary} assumes no structure  of the measurement.
However, there are structures that can be exploited in order to reduce the number of free parameters in the unitary operators.
In the scenario where $k$ copies of the state is used, we note the global state is permutationally invariant over the tensor structure. 
For example, if we take two copies of our two-qubit system, we can write the 4-qubit Hamiltonian as
\begin{equation}
H = \sum_{i,j,k,l=1}^{4} \alpha_{i,j,k,l} \sigma_i \otimes \sigma_j \otimes \sigma_k \otimes  \sigma_l.
\end{equation}
This gives us permutational invariance over the 2-qubit level, that is, 
\begin{equation}
\alpha_{i,j,k,l} = \alpha_{k,l,i,j}.
\end{equation}
This invariance does not hold at the $i$-$j$ level as we do not require the single copy states to be permutationally invariant (although for our problem we do find numerically the permutational-invariant space to contains the optimal states.
This is an agreement with the consensus for noise models other than amplitude damping~\cite{Frowis2014,Koczor2019}).
In this way we restrict our projectors to also being permutationally invariant, which allows us to work with fewer parameters to optimise.

The number of parameters is the character of $\text{SU}(2^{qk})/\text{SU}(2^q)^{\otimes k}$. The dimension of this Hilbert space for $q$-qubits and $k$-copies is
\begin{equation}
\text{dim} \mathcal{H}(q,k) =  \frac{1}{(4^q-1)!} \Pi_{i=1}^{4^q-1}(k+i).
\end{equation}
This is a reduction from an exponential in $q$ and $k$ to a polynomial of order $4^q-1$ in $k$. 

\subsection{\label{subsubsec:PSO}Particle Swarm Optimisation}

Particle Swarm Optimisation (PSO)~\cite{pyswarmsJOSS2018} is a global optimiser which takes multiple particles (points in the input space) and moves them through the search space given a local and global current best known answer~\cite{KE95}. This gives us a sense of position and velocity for each particle. Over each iteration a particle's position is updated by a velocity which is weighted between the local and global currently known best solution. PSO is intended to mimic the collective behaviour of animals such as schooling fish or ants, where the animals are free to move along their own path but within the confines of some collective behaviour. 

Let $x_i \in \mathbb{R}^m$ be the $i^{\text{th}}$ particle's position, with velocity  $v_i \in \mathbb{R}^m$. We also need to track to current best known position $y_i$ for the $i^{\text{th}}$ particle along with $\tilde{y}$, the global best known position. We also have three meta-parameters $\omega, c_1, c_2$, which are the inertia factor, local solution bias and global solution bias respectively. Each position  $x_i$ is updated as per
\begin{align}
x_{i}(t+1) &= x_{i}(t) + v_{i}(t+1),\\
v_{i}(t + 1) &= \omega * v_{i}(t) + c_{1}[y_{i}(t) - x_{i}(t)] + c_{2}[\tilde{y}(t) - x_{i}(t)],
\end{align}
and the optimal solutions are updated if needs be. This procedure is repeated until a termination condition is met. Here termination is the maximum number of iterations we can computationally afford. This is dependant on the optimal result having not changed for the last set of iterations (typically we choose to check the last 100) to ensure the search has been exhausted.

\section{\label{subsec:Qcircuit}Quantum circuit optimisation}

In this Appendix, we describe the type of programmable quantum circuit employed in this work and the algorithm for optimising them.
The quantum circuits are used to decompose the unitary operators necessary to attain the HCRB.
However, instead of finding the unitaries and then decomposing the unitaries into circuits, we instead replace the unitaries with quantum circuits and then optimise the circuit parameters using one of the two optimisers, the genetic-gradient algorithm or the differential evolution (DE). In order to terminate this optimisation we run until at least the cost function is within some small tolerance of the equivalent unitary optimisation (Sec.~\ref{subsec:numerical_results}) result and then allow it to run for a few more iterations to ensure minimisation of gates too. 

\subsection{\label{subsubsec:VQC}Programmable quantum circuit}
\begin{figure}[h]
	\centering
	\hspace{15pt}
	\Qcircuit @C=1em @R=.7em {
		&\lstick{\ket{0}}  & \gate{R_{1,1} }  & \targ     & \gate{R_{2,1}} &  \ctrl{1} & \gate{R_{3,1}} & \multigate{1}{U(\bm{\varphi})} & \gate{R_{4,1} }  & \targ     & \gate{R_{5,1}} &  \ctrl{1} & \gate{R_{6,1}} & \qw & \meter \\
		&\lstick{\ket{0}}  & \gate{R_{1,2} }  & \ctrl{-1} & \gate{R_{2,2}} &  \targ     & \gate{R_{3,2}} & \ghost{U(\vec{\varphi})} & \gate{R_{4,2} }  & \ctrl{-1} & \gate{R_{5,2}} &  \targ     & \gate{R_{6,2}} & \meter\\
		& & & & & & & & & & & & & \dstick{C_1} \cwx & \dstick{C_2} \cwx[-2] \gategroup{1}{3}{2}{7}{.7em}{--} \gategroup{1}{9}{2}{13}{.7em}{--}
	}
	\caption{General two-qubit circuit with a single gadget for 3D magnetometry.}
	\label{FIG:vqc}
\end{figure}
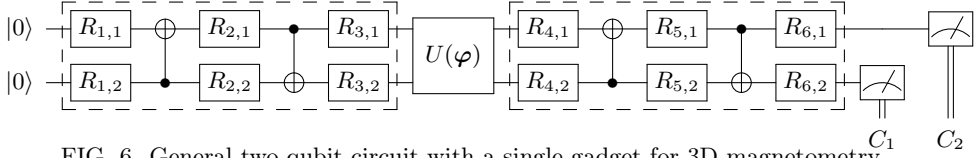

The quantum circuit that we consider for this work~\cite{Benicio} consisted of a sequence of units we call gadgets. Each gadget for $n$ qubits consisted of $n+1$ layers of rotational single-qubit gates \begin{equation}
R_{i,j} = \exp\left(-\text{i}(\alpha_{i,j,x}\sigma_x+\alpha_{i,j,y}\sigma_y+\alpha_{i,j,z}\sigma_z )\right),
\end{equation}
where $i$ is the layer and $j$ is the qubit identifier.
In between these layers are a series of CNOT gates connected to a single control qubit and controlling all the other qubits. After the next layer of single-qubit gates we move the control qubit down one. This gives us $n(n+1)$ single-qubit gates and $n(n-1)$ two-qubit gates. An example of single gadget circuits for two qubits is shown in~Fig.~\ref{FIG:vqc}.

We break down the circuit optimisation problem into a discrete part and a continuous part.
The discrete part consists of the on-off switch to the CNOT and rotation gates, whereas the continuous part tunes the $\hat{\alpha}_{i,j}$ for all the continuous gates.
The discrete component is parameterised by two bit strings, one representing the single-qubit gates and the other the two-qubit gates. 
If a bit in the string is 1, the gate is ``on'' and if a bit is 0, the gate is ``off''.
While it is true that there is no need to have the single-qubit gates with on-off states as we could set all the relevant bits to 1 and get the same effect, we choose to leave that option in. This is to minimise the total number of gates and lessen the numerical burden.
To lessen the numerical burden further, we initialise the circuit with one gadget in each circuit and include a probability of adding more gadgets to the optimisation algorithms. This approach minimises the number of CNOT gates used, which is typically the harder gate to implement in experiments.

\subsection{\label{subsubsec:genetic alg.}Genetic-gradient piecewise algorithm}

\begin{figure}[t]
	\begin{center}
		\begin{tikzpicture}[node distance = 0.5cm, auto, scale = 0.75, every node/.style={scale=.75}]
		\node [block] (init) {Initialise population};
		\node [block, right =  of init] (eval) {Evaluate fitness};
		\node [cloud, below = 0.5cm of eval] (grad) {Optimise  single qubit};
		\node [block, right = of  eval] (select) {Randomly select new population, ($\propto$ fitness)};
		\node [block, right = of  select] (cross) {Cross-over parents to generate children};
		\node [block, right = of  cross] (mutate) {Mutate children (random bit-flip)};
		\node [block, right = of mutate] (eval_chil) {Evaluate fitness of children};
		\node [block, right = of  eval_chil] (new_pop) {Generate new pop; select best from child and parents};
		\node [decision, below = 0.5cm of new_pop] (term) {termination condition reached?};
		\node [block, below = of term] (end) {return optimal solution};
		\path [line] (init) -- (eval);
		\path [line] (eval) -- (select);
		\path [line, dashed] (grad) -- (eval);
		\path [line] (select) -- (cross);
		\path [line] (cross) -- (mutate);
		\path [line] (mutate) -- (eval_chil);
		\path [line] (eval_chil) -- (new_pop);
		\path [line] (new_pop) -- (term);
		\path [line] (term) -| node [near start]{no} (select);
		\path [line] (term) -- node {yes} (end);
		\path [line, dashed] (eval_chil) |- (grad);
		\end{tikzpicture}
	\end{center}
	\caption{Flowchart for genetic algorithm.}
	\label{FIG:gen_alg}
\end{figure}
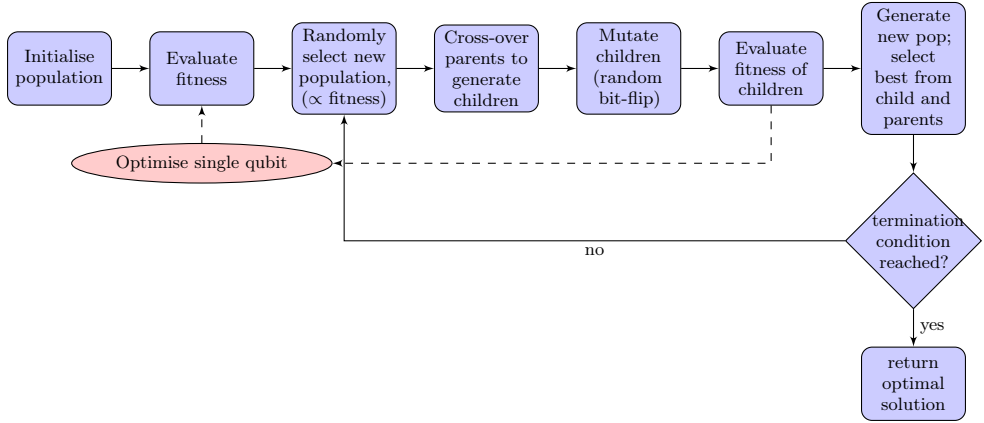

This optimiser was historically developed for optimising engineering structures  by combining the traditional genetic algorithm~\cite{Wu1995} for discrete optimisation and a gradient-based algorithm~\cite[Chap. 9]{boyd2004} for continuous optimisation.
However, the algorithm is capable of working with any of the parameterisations, and the flowchart for the algorithm used can be seen in Figure \ref{FIG:gen_alg}.
It follows largely a typical genetic algorithm with genetic elitism, except  the evaluation of the cost function involves the optimisation of the continuous single-qubit gate parameters.
This can, of course, be done with any continuous optimisation technique and in this work we have found a simple gradient descent to perform well. We use the result of the unitary optimisation as a sufficient termination condition and allow the algorithm to continue for remaining wall-time once the unitary optimisation result has been obtained. 

\subsection{\label{subsubsec:DE}Differential Evolution}

Another optimiser we employ to optimise the quantum circuit is differential evolution (DE)~\cite{Storn1997}.
Originally, designed for non-convex continuous optimisation, we modify the algorithm slightly to accommodate discrete optimisation too.
In this instance, the switch is used to turn a layer of the circuit instead of a gate on and off, which reduces the number of switches that DE needs to search from $2n^2$ down to $2n+1$, where $n$ is the total number of qubits of the system.

In our work, the DE algorithm searches for a solution that contains two parts---a discrete part which is a sequence of bits that is $2n+1$ long, and a continuous part that is a sequence of $3n(n+1)$ numbers in $\left[0,2\pi\right)$. The algorithm works as follows.
\begin{algorithm}[H]
	\caption{Pseudocode for DE including discrete optimisation}
	\label{alg:DE}
	\begin{algorithmic}[1]
		\Require population size $N_p$, largest iteration time $T$, scaling factor $\mathcal{F}$, and the crossover rate $C_r$
		\Ensure Solution with lowest $\CC$
		\State{Initialise population of random solution candidates, each with discrete and continuous part}
		\ForAll \State~{Compute $\CC$} \EndFor
		\While {The terminating condition is not yet reached}
		\ForAll \State~{Select 3 members from population at random} \State~{Create offspring} 
		\State~{Compute offspring's $\CC$}
		\If {$\CC_{\text{offspring}}<\CC_{\text{candidate}}$} \State~{Replace candidate with offspring} \EndIf
		\EndFor
		\EndWhile
	\end{algorithmic}
\end{algorithm}
A population of $N_p$ solution candidates is initialised and the $\CC$ is computed for each. The offsprings are then computed for each triple selected from the population. For the continuous part of the solution, we construct the offspring using
\begin{equation}
\label{eq:Offspring_continuous}
D_{i}(t)^{(j)}=\begin{cases}
V_{i,1}(t)^{(j)}+ \mathcal{F}\cdot(V_{i,2}(t)^{(j)}-V_{i,3}(t)^{(j)}) & \text{if}~\operatorname{rand(0,1)}\leq C_r\\
V_{i}(t)^{(j)} & \text{otherwise},\\
\end{cases}
\end{equation}
where $D_i$ is the offspring for the $i^{\text{th}}$ candidate. The index $j$ indicate the dimension of the search space, and $t$ indicate the iteration time step.
As for the discrete part of the solution, a similar equation is used but with $\mathcal{F}=1$, 
and the modulo of $D_{i}(t)^{(j)}/2$ is computed at the end.
\begin{equation}
\label{eq:Offspring_discrete}
D_{i}(t)^{(j)}=\begin{cases}
\left(V_{i,1}(t)^{(j)}+ (V_{i,2}(t)^{(j)}-V_{i,3}(t)^{(j)})\right) \operatorname{mod} 2 & \text{if} \operatorname{rand(0,1)}\leq C_r\\
V_{i}(t)^{(j)} & \text{otherwise}.
\end{cases}
\end{equation}
The $\CC$ of the offspring is then computed and if this $\CC$ is smaller than the $\CC$ of the candidate, the offspring replaces the candidate for the next time step $t+1$. Rand(0,1) is a random number chosen uniformly in the interval $[0,1]$.
Our algorithm accepts the solution after $T$ iterations of the algorithms has passed.

The DE algorithm is tuned using 4 parameters: population size $N_p$, largest iteration time $T$, scaling factor $\mathcal{F}$, and the crossover rate $C_r$.
For this work, we set $N_p=150$, $T=600$, $F=0.9$, and $C_r=0.3$, which are found through trial and error by investigating the variance of population as a function of time. The idea is to maintain the exploration power of the algorithm for as long as possible in order to increase the possibility of attaining a globally optimum solution.

\begin{figure}[t]
	\centering
	\includegraphics[width=0.5\textwidth]{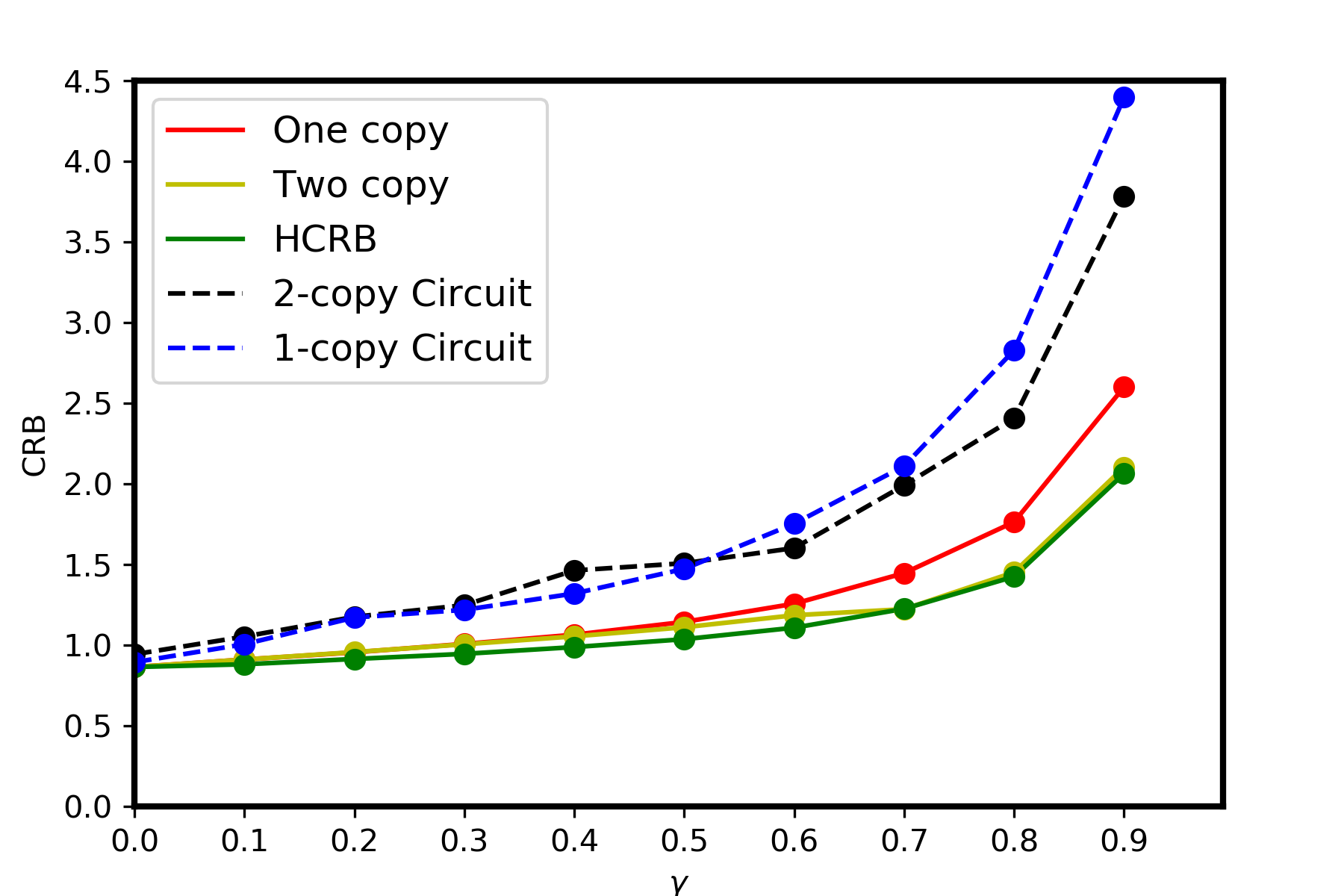}
	\caption{The $\CC$ from optimised quantum circuits using DE. for two-copy (black dashed line) and one-copy measurement (blue dashed line).
	The solid lines are those from Fig.~\ref{FIG:opt_state_and_measure} obtained by optimising unitaries for the one-copy (red solid line), two-copy (light green solid line), and the HCRB (green solid line).}
	\label{fig:circuitCFI}
\end{figure}

The $\CC$ from the genetic-inspired algorithm is able to reach the unitary's $\CC$ and so used the value from the unitary as the termination condition whereas the $\CC$ from the DE is unable to surpass the one-copy $\CC$ set by the unitary optimisation as seen in Fig~\ref{fig:circuitCFI}.
However, adding the second copy of the input state does show an advantage in the high noise regime ($\gamma > 0.5$) just similar to what has been observed in unitary optimisation.

The difference in performance between the two optimisers is most likely due to its handling of the discrete optimisation, i.e., the on-off switch of the gate. 
This conclusion is drawn from two main observations.
Firstly,  both the optimisers employ the same quantum circuit representations but differ in how the on-off switch is handled.
Secondly, DE is a global optimiser for continuous space and hence is capable, in principle, of reaching the same solution in the space as a gradient-based optimiser used in the genetic-inspired algorithm. However, the performance of the genetic-inspired algorithm far exceeds the DE algorithm even as sufficient computational time is given for the DE to converge, which suggests that the continuous search space is not the part preventing DE from reaching a near optimal solution. We investigate this difference further by looking into the circuit solution in Appendix~\ref{app:Optimised quantum circuits}.

\subsubsection{\label{app:Optimised quantum circuits} Optimised quantum circuits}

Here we present the results from optimising the quantum circuit using the DE algorithm in Appendix~\ref{subsubsec:DE}.
The circuits obtained for two-copies of two-qubit 3D magnetometry are shallow (Fig.~\ref{fig:DE_circuits_2copy}). Unlike the circuits in Fig.~\ref{fig:circ} obtained from the genetic-inspired algorithm, we clearly see why these circuits do not reach the unitary bounds. The DE algorithm appears to favour circuits with reduced entanglement in the case of medium and high noise. Fig.~\ref{fig:DE_circuits_2copy}(b) and Fig.~\ref{fig:DE_circuits_2copy}(c) both lack the CNOT gates either in the input-state generation or in the measurement stage, which is not observed in the circuits found by the genetic-inspired algorithm (Fig.~\ref{fig:circ}) although DE was not explicitly programmed to minimise the number of CNOT gates.
\begin{figure}[t]

	\centering

\begin{minipage}{\linewidth}
\centering
\hspace{20pt}
	\Qcircuit @C=1em @R=.7em {
		\lstick{\ket{0}} &\gate{R_1^{(1)}} & \targ&\gate{R_2^{(1)}} & \qw &
		\multigate{1}{U_{\vec{\varphi}}}& \qw &\gate{R_3^{(1)}}
		&\targ&\qw&\qw&\targ&\qw&\qw 
		&\gate{R_4^{(1)}} & \qw & \measureD{z}
		\\
		\lstick{\ket{0}} &\gate{R_1^{(2)}} &\ctrl{-1} &\gate{R_2^{(2)}} & \qw &
		\ghost{U_{\vec{\varphi}}}&
		\qw &\gate{R_3^{(2)}}
		&\qw&\targ&\qw&\qw&\targ&\qw
		&\gate{R_4^{(2)}} & \qw & \measureD{z}
		\\
		\lstick{\ket{0}} &\gate{R_1^{(1)}} & \targ&\gate{R_2^{(1)}} & \qw &
		\multigate{1}{U_{\vec{\varphi}}}& \qw &\gate{R_3^{(1)}}
		&\ctrl{-2}&\ctrl{-1}&\ctrl{1}&\qw&\qw&\targ
		&\gate{R_4^{(1)}} & \qw & \measureD{z}
		\\
		\lstick{\ket{0}} &\gate{R_1^{(2)}} &\ctrl{-1} &\gate{R_2^{(2)}} & \qw &
		\ghost{U_{\vec{\varphi}}}& \qw 
		&\gate{R_3^{(2)}}
		&\qw&\qw&\targ&\ctrl{-3}&\ctrl{-2}&\ctrl{-1}
		&\gate{R_4^{(2)}} & \qw & \measureD{z}	
	}~\\~\\
	(a) Low noise, two copies 
	\end{minipage}

 	~~\\
	~~\\

	\begin{minipage}{.45\linewidth}
	\centering
\Qcircuit @C=1em @R=.7em {
		\lstick{\ket{0}} &\gate{R_1^{(1)}} & \targ&\gate{R_2^{(1)}} & \ctrl{1} & \qw &
		\multigate{1}{U_{\vec{\varphi}}}& \qw &\gate{R_3^{(1)}}
		& \qw & \measureD{z}
		\\
		\lstick{\ket{0}} &\gate{R_1^{(2)}} &\ctrl{-1} &\gate{R_2^{(2)}} &\targ& \qw &
		\ghost{U_{\vec{\varphi}}}&
		\qw &\gate{R_3^{(2)}}
		& \qw & \measureD{z}
		\\
		\lstick{\ket{0}} &\gate{R_1^{(1)}} & \targ&\gate{R_2^{(1)}} & \ctrl{1}& \qw &
		\multigate{1}{U_{\vec{\varphi}}}& \qw &\gate{R_3^{(1)}}
		& \qw & \measureD{z}
		\\
		\lstick{\ket{0}} &\gate{R_1^{(2)}} &\ctrl{-1} &\gate{R_2^{(2)}} &\targ& \qw &
		\ghost{U_{\vec{\varphi}}}& \qw 
		&\gate{R_3^{(2)}}
		& \qw & \measureD{z}	}~\\
	
	(b) Medium noise, two copies 
	\end{minipage}
	~
 	\hspace{5pt}
	\begin{minipage}{.45\linewidth}
 	\centering
	\Qcircuit @C=1em @R=.7em {
		\lstick{\ket{0}} &\gate{R_1^{(1)}} & \qw &
		\multigate{1}{U_{\vec{\varphi}}}& \qw &\gate{R_3^{(1)}}
		&\targ&\qw&\qw
		& \qw & \measureD{z}
		\\
		\lstick{\ket{0}} &\gate{R_1^{(2)}} & \qw &
		\ghost{U_{\vec{\varphi}}}&
		\qw &\gate{R_3^{(2)}}
		&\qw&\targ&\qw
		& \qw & \measureD{z}
		\\
		\lstick{\ket{0}} &\gate{R_1^{(1)}} &  \qw &
		\multigate{1}{U_{\vec{\varphi}}}& \qw &\gate{R_3^{(1)}}
		&\qw&\qw&\targ
		& \qw & \measureD{z}
		\\
		\lstick{\ket{0}} &\gate{R_1^{(2)}}& \qw &
		\ghost{U_{\vec{\varphi}}}& \qw 
		&\gate{R_3^{(2)}}
		&\ctrl{-3}&\ctrl{-2}&\ctrl{-1}
		& \qw & \measureD{z}	
	}~\\
	(c) High noise, two copies
	\end{minipage}
	\caption{Quantum circuits generated using DE for two-copies of two-qubit 3D magnetometry for (a) $\gamma=0$ (b) $\gamma=0.5$ (c) $\gamma = 0.9$}
	\label{fig:DE_circuits_2copy}
\end{figure}

\end{document}